\documentclass[%
reprint,
superscriptaddress,
amsmath,amssymb,
aps,
prd,
floatfix,
]{revtex4-2}

\usepackage{graphicx}% Include figure files
\graphicspath{{./figures/}}
\usepackage{xcolor}
\usepackage{ulem}
\usepackage{dcolumn}% Align table columns on decimal point
\usepackage{bm}% bold math
\usepackage{hyperref}
\hypersetup{colorlinks, linkcolor = [rgb]{0, 0, 0.5}, citecolor = [rgb]{0,0.0,0.5}, urlcolor = [rgb]{0,0.0,0.5}}
\usepackage[caption=false]{subfig}
\usepackage{siunitx}
\usepackage[capitalise]{cleveref}
\allowdisplaybreaks

\begin{document}

\title{Extraction of trans-helicity worm-gear distributions and opportunities at the Electron-Ion Collider in China}

\newcommand*{\PKU}{School of Physics, Peking University, Beijing 100871, China}\affiliation{\PKU}
\newcommand*{\CHEP}{Center for High Energy Physics, Peking University, Beijing 100871, China}
\newcommand*{\CICQM}{Collaborative Innovation Center of Quantum Matter, Beijing, China}
\newcommand*{\SDU}{Key Laboratory of Particle Physics and Particle Irradiation (MOE), Institute of Frontier and Interdisciplinary Science, Shandong University, Qingdao, Shandong 266237, China}\affiliation{\SDU}
\newcommand*{\SCNT}{Southern Center for Nuclear-Science Theory (SCNT), Institute of Modern Physics, Chinese Academy of Sciences, Huizhou 516000, China}
\newcommand*{\IMP}{Institute of Modern Physics, Chinese Academy of Sciences, Lanzhou, Gansu Province 730000, China}
\newcommand*{\NNU}{Department of Physics and Institute of Theoretical Physics,
Nanjing Normal University, Nanjing, Jiangsu 210023, China}
\newcommand*{\UCAS}{University of Chinese Academy of Sciences, Beijing 100049, China}
\newcommand*{\KLQLPIPP}{Key Laboratory of Quark and Lepton Physics (MOE) and Institute of Particle Physics, Central China Normal University, Wuhan 430079, China}

\author{Ke~Yang}\email{yangke2020@stu.pku.edu.cn}\affiliation{\PKU}
\author{Tianbo~Liu}\email{liutb@sdu.edu.cn}\affiliation{\SDU}\affiliation{\SCNT}
\author{Peng~Sun}\email{pengsun@impcas.ac.cn}\affiliation{\IMP}\affiliation{\UCAS}
\author{Yuxiang~Zhao}\email{yxzhao@impcas.ac.cn}\affiliation{\SCNT}\affiliation{\IMP}\affiliation{\UCAS}\affiliation{\KLQLPIPP}
\author{Bo-Qiang~Ma}\email{mabq@pku.edu.cn}\affiliation{\PKU}\affiliation{\CHEP}\affiliation{\CICQM}

\begin{abstract}

We present a global analysis of the trans-helicity worm-gear distribution function, $g_{1T}^\perp$, by fitting the longitudinal-transverse double spin asymmetry data of the semi-inclusive deep inelastic scattering. The analysis is performed within the framework of transverse momentum dependent factorization and evolution. It is found that the $u$-quark favors a positive distribution and the $d$-quark favors a negative distribution, which is consistent with previous model calculations and phenomenological extractions. Based on the fit to existing world data, we also study the impact of the proposed electron-ion collider in China and conclude that it can significantly improve the precision of the worm-gear distribution function and hence enhance our understanding of nucleon spin structures. 

\end{abstract}

\maketitle

\section{Introduction}

Understanding the internal structure of nucleons is pivotal for comprehending the strong force that binds quarks and gluons within nucleons, and for shedding light on the fundamental properties of the matter. 
In recent years, the pursuit of multi-dimensional tomography of 
the nucleon has emerged as a cutting-edge approach to probe distributions of quarks and gluons within the 
nucleon, offering a deeper understanding of its internal dynamics. Transverse momentum dependent (TMD) parton distribution functions (PDFs) contain the information of the parton transverse momentum with respect to the parent nucleon, and hence provide three-dimensional imaging of the nucleon in the momentum space.

At the leading twist, there are eight TMDs for quarks~\cite{Tangerman:1994eh,Mulders:1995dh,Boer:1997nt,Ji:2002aa}.
Among them, the worm-gear-T distribution $g_{1T}^\perp(x,k_T^2)$, also known as the trans-helicity distribution~\cite{Zhu:2011ym,Zhu:2011zza} or the Kotzinian-Mulders function~\cite{LuoXuan:2020icc,Kotzinian:1995cz}, describes the probability density of finding a longitudinally polarized quark with longitudinal momentum fraction $x$ and transverse momentum $k_T$ in a transversely polarized nucleon. As well as the worm-gear-L, or longi-transversity, distribution $h_{1L}^\perp$ that describes the probability density of finding a transversely polarized quark in a longitudinally polarized nucleon, it can be expressed as the overlap between wave functions differing by one unit of orbital angular momentum~\cite{Miller:2007ae,Bacchetta:2008af,Pasquini:2008ax,Pasquini:2009bv,Burkardt:2007rv}, and many efforts have been devoted to the worm-gear TMDs to understand nucleon spin and flavor structures. 

Although the two worm-gear distributions are defined as independent quantities from the decomposition of the quark-quark correlator, some relation, such as $g_{1T}^\perp=-h_{1L}^{\perp}$, is suggested based on quark model like calculations~\cite{Jakob:1997wg,Pasquini:2008ax,Efremov:2009ze,Avakian:2010br,Zhu:2011ym,Zhu:2011zza}.
Following the SU(6) spin-flavor structure, the $g_{1T}^\perp$ distribution of the up quark was predicted to be positive and with a greater magnitude than the negative down quark distribution, and explicit calculations have been done in the light-cone constituent quark model~\cite{Bacchetta:1999kz,Ji:2002xn,Pasquini:2008ax,Boffi:2009sh,Bacchetta:2010si}, the spectator diquark model~\cite{Bacchetta:2010si,Jakob:1997wg,Gamberg:2007wm,Bacchetta:2008af}, the MIT bag model ~\cite{Avakian:2010br}, and the covariant parton model~\cite{Efremov:2009ze}. 
On the other hand, the large-$N_c$ approximation~\cite{Pobylitsa:2003ty} states that the worm-gear distributions of up quark and down quark only  differ by a sign and have the same magnitude, {\it i.e.} $g_{1T}^{\perp u} = -g_{1T}^{\perp d}$. Besides, if taking the Wandzura-Wilczek (WW)-type approximation~\cite{Avakian:2007mv,Accardi:2009au,Kanazawa:2015ajw,Scimemi:2018mmi,LuoXuan:2020icc,Kotzinian:1995cz},
which neglects the contribution from quark-gluon-quark correlations, one may relate the trans-helicity worm-gear distribution to the helicity distribution as 
\begin{equation}
g_{1 T}^{\perp(1)}(x) 
 \stackrel{\text {WW}}{\approx} x \int_x^1 \frac{d y}{y} g_1(y),  
\end{equation}
where
\begin{equation}
g_{1 T}^{\perp(1)}(x)  \equiv  \pi\int d^2 {k}_{T} \frac{k_{T}^2}{2 M^2} g_{1 T}^{\perp}(x, {k}_{T}^2),
\end{equation}
is the first transverse moment that has also been studied in lattice QCD~\cite{Hagler:2009mb,Musch:2010ka,Yoon:2017qzo}.

In experiment, the semi-inclusive deep inelastic scattering (SIDIS) is one of the main processes to study TMDs. According to the TMD factorization, the trans-helicity distribution $g_{1T}^{\perp}$ contributes to a double spin asymmetry $A_{LT}$ with azimuthal modulation as $\cos(\phi_{h} - \phi_{S})$. With the development of polarized beams and targets, this asymmetry has been measured by HERMES~\cite{HERMES:2020ifk}, COMPASS~\cite{COMPASS:2016led,Parsamyan:2018evv,Avakian:2019drf} and Jefferson Lab (JLab)~\cite{JeffersonLabHallA:2011vwy}. In some recent phenomenological analyses~\cite{Bhattacharya:2021twu,Horstmann:2022xkk}, it was found that the extracted worm-gear distributions supported the positive result for the up quark and the negative result for the down quark as suggested by the model calculations. However, due to the limited accuracy of existing world data, one has to introduce some bias in the fit to obtain reasonable results and almost no constraint is put on sea quarks.

The Electron-Ion Collider in China (EicC) is proposed as a future facility in nuclear physics, and one of its main physics goals is to precisely measure nucleon TMDs via the SIDIS process. It is designed to deliver a $3.5\,\rm GeV$ electron beam with 80\% polarization colliding with various types of ion beams. The designed energy of the proton beam is $20\,\rm GeV$ and correspondingly the energy of the $^{3}$He beam is $40\,\rm GeV$. Both the proton and the $^{3}$He beams can be longitudinally or transversely polarized with 70\% polarization. The instantaneous luminosity can reach about $2\times10^{33}\,\mathrm{cm}^{-2}\mathrm{s}^{-1}$. The EicC kinematic coverage will fill the gap between multi-hall SIDIS program at the $12\,\rm GeV$ upgraded JLab, which covers relatively large-$x$ region dominated by valence quarks, and the Electron-Ion Collider (EIC) to be built at the Brookhaven National Laboratory (BNL), which can reach the small-$x$ region down to about $10^{-4}$~\cite{Accardi:2012qut,AbdulKhalek:2021gbh}.
Therefore, a combination of all these facilities is expected to provide precise determination of TMDs in a full kinematic coverage~\cite{Anderle:2021dpv}, towards a complete three-dimensional imaging of nucleon spin structures.

In this paper, we perform a global analysis of trans-helicity TMDs by fitting the longitudinal-transverse double spin asymmetry data from HERMES, COMPASS and JLab.
Taking the world data fit result as the baseline, we further study the impact of the EicC SIDIS program on the determination of the worm-gear distribution. 
The rest of the paper is organized as follows. 
In Sec.~\ref{Sec II}, we briefly review the theoretical framework. 
In Sec.~\ref{Sec III}, we present the parametrization of the trans-helicity worm-gear distributions and the fit results to world data.
In Sec.~\ref{Sec III.2}, we study the EicC impact on the extraction of the worm-gear distributions by adding simulated pseudodata in the fit.
A summary is drawn in Sec.~\ref{Sec IV}. 

\section{Theoretical formalism}
\label{Sec II}

We consider the SIDIS process
\begin{equation}
\ell(l)+N(P)\longrightarrow\ell(l^{\prime})+h(P_{h})+X,
\end{equation}
where $\ell$ represents the lepton, $N$ represents the nucleon, and $h$ represents the detected hadron. The four-momenta of corresponding particles are given in parentheses. The commonly used kinematic variables for the SIDIS process are defined as
\begin{gather}
Q^2 = -(l-l^\prime)^2 = -q^2,\\
x=\frac{Q^2}{2P\cdot q},
\quad 
y=\frac{P\cdot q}{P\cdot l},
\quad
z=\frac{P\cdot P_{h}}{P\cdot q},\\
\gamma = \frac{2xM}{Q} = \frac{M Q}{P\cdot q},
\end{gather}
where $q = l - l'$ is the transferred momentum and $M$ is the nucleon mass.

For the SIDIS process with a transversely polarized target and a longitudinally polarized lepton beam, one can write the differential cross section within the one-photon-exchange approximation as~\cite{Bacchetta:2006tn}
\begin{equation}
 \begin{aligned}
	&\frac{d\sigma}{d x d y  d z d \phi_{h} d \phi_{S} d P_{hT}^2} =  
    \sigma_0 \Big\{F_{UU} \\
    &\quad 
    + \lambda_{e} | S_{\perp} | 
    \Big[ 
    \sqrt{1-\varepsilon^2}
    \cos(\phi_h  -\phi_S)
    F_{LT}^{\cos(\phi_h  -\phi_S)} \\
    &\quad 
    + \sqrt{2\varepsilon(1-\varepsilon)}
    \cos(2\phi_h  -\phi_S)
    F_{LT}^{\cos(2\phi_h  -\phi_S)} \\
    &\quad
    + \sqrt{2\varepsilon(1-\varepsilon)}
    \cos(\phi_S)
    F_{LT}^{\cos\phi_S} 
    +\cdots\Big] \Big\},
		\label{eq:CSLT}
  \end{aligned}
\end{equation}
where
\begin{equation}
 \sigma_0 = \frac{\alpha^2}{x y Q^2}\frac{y^2}{2(1-\varepsilon)}\left(1+\frac{\gamma^2}{2x}\right),
\end{equation}
$\alpha$ is the electromagnetic fine structure constant, $|S_{\perp}|$ represents the transversal component of the nucleon spin vector, $\lambda_e$ represents the helicity of the lepton beam, and $\varepsilon$ is the ratio of longitudinal and transverse photon flux,
\begin{equation}
\varepsilon = \frac{1-y-\frac{1}{4}\gamma^2 y^2}{1-y+\frac{1}{2}y^2 +\frac{1}{4}\gamma^2 y^2}.
\end{equation}
\begin{figure}[htbp]
    \centering
    \includegraphics[scale=0.16]{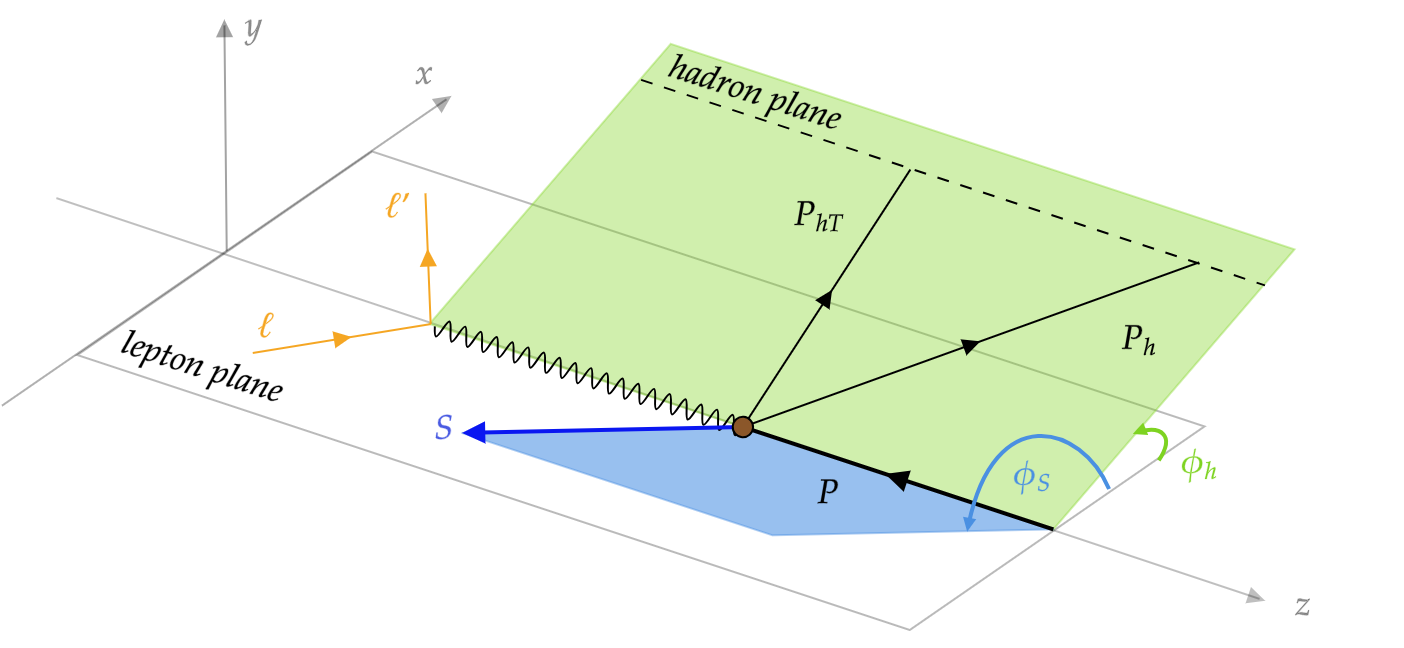}
    \caption{Trento conventions of the transverse momentum and azimuthal angles.}
    \label{Fig:1}
\end{figure}

As shown in Fig. \ref{Fig:1}, we follow the Trento conventions~\cite{Bacchetta:2004jz}, in which the momenta of the virtual photon and the nucleon are chosen along the $\hat{z}$ direction. 
One can express the transverse momentum $P_{hT}$ of the hadron and the azimuthal angles $\phi_h$ and $\phi_S$ in Lorentz invariant forms as
\begin{gather}
P_{hT}=\sqrt{-g^{\mu\nu}_{\bot}P_{h\mu}P_{h\nu}},\\
\cos \phi_h = -\frac{l_\mu P_{h\nu}g^{\mu\nu}_{\bot}}{l_{\bot}P_{hT}}, 
\quad
\sin \phi_h = -\frac{l_\mu P_{h\nu} \epsilon^{\mu\nu}_{\bot}}{l_{\bot}P_{hT}}, \\
\cos \phi_S = -\frac{l_\mu S_{\bot\nu}g^{\mu\nu}_{\bot}}{l_{\bot}S_{\bot}},
\quad
\sin \phi_S = -\frac{l_\mu S_{\bot\nu} \epsilon^{\mu\nu}_{\bot}}{l_{\bot}S_{\bot}},
\end{gather}
where $l_{\bot} = \sqrt{-g^{\mu\nu}_{\bot} l_\mu l_\nu}$ and $S_{\bot} = \sqrt{-g^{\mu\nu}_{\bot} S_\mu S_\nu}$ with $S^\mu$ being the spin vector of the nucleon. The transverse metric and the transverse antisymmetry tensor are defined as
\begin{gather}
g^{\mu\nu}_{\bot} = g^{\mu\nu} - \frac{q^\mu P^\nu + P^\mu q^\nu}{P\cdot q (1 + \gamma^2)} + \frac{\gamma^2}{1+\gamma^2}\left(\frac{q^\mu q^\nu}{Q^2} - \frac{P^\mu P^\nu}{M^2} \right),\\
\epsilon^{\mu\nu}_{\bot} = \epsilon^{\mu\nu\rho\sigma} \frac{P_{\rho} q_{\sigma}}{P\cdot q\sqrt{1+\gamma^2}},
\end{gather}
where $\epsilon^{\mu\nu\rho\sigma}$ is the totally antisymmetric tensor with the convention $\epsilon^{0123} = 1$.

The worm-gear distribution $g_{1T}^\perp$ can be extracted from the longitudinal-transverse double spin asymmetry, which is given by the ratio between the structure functions $F_{LT}^{\cos(\phi_h  -\phi_S)}$ and $F_{UU}$.
According to the TMD factorization~\cite {Bacchetta:2006tn}, the structure functions at low transverse momentum, {\it i.e.} small $\delta = |P_{hT}|/(z Q)$, can be approximated in terms of TMD PDF and TMD fragmentation function (FF) as
\begin{align}
&F_{U U}=\big|C_V\big(Q^2, \mu\big)\big|^2 
x\sum_q e_q^2 \int_0^{\infty} \frac{b_{T} db_{T}}{2 \pi} 
J_0\left(\frac{b_{T}P_{hT}}{z}\right)
\notag\\
&\quad  \times f_{1, q \leftarrow H}(x, b_{T} ; \mu, \zeta) D_{1, q \rightarrow h}(z, b_{T} ; \mu, \bar{\zeta})+\mathcal{O}\left(\frac{P_{hT}^2}{Q^2}\right), 
\\
&F_{L T}^{\cos \left(\phi_h-\phi_S\right)}=\left|C_V\left(Q^2, \mu\right)\right|^2 x\sum_q e_q^2 M \int_0^{\infty} \frac{b_{T}^2 db_{T}}{2 \pi}
\notag\\
& \quad \times J_1\left(\frac{b_{T}P_{hT}}{z}\right) g_{1 T, q \leftarrow H}^\perp(x, b_{T} ; \mu, \zeta) D_{1, q \rightarrow h}(z, b_{T} ; \mu, \bar{\zeta})
\notag\\
&\quad +\mathcal{O}\left(\frac{P_{hT}^2}{Q^2}\right), 
\end{align}
where $e_q$ is the electric charge of the quark with flavor $q$, $C_V$ is the hard factor that can be calculated via perturbative QCD, and $J_0$ and $J_1$ are the first kind Bessel functions. Here the unpolarized TMD PDF $f_1$, the worm-gear TMD PDF $g_{1T}^\perp$, and the unpolarized TMD FF $D_1$ are given in $b_T$ space. They are related to corresponding functions in the transverse momentum space through Fourier transforms,
\begin{equation}
f_1(x,k_{T};\mu,\zeta) = \int_0^\infty \frac{b_{T}db_{T}}{2\pi} J_0(b_{T} k_{T}) f_1(x,b_{T};\mu,\zeta),    
\end{equation}
\begin{equation}
\begin{aligned}
\frac{k_{T}}{M}g_{1T}^\perp & (x,k_{T};\mu,\zeta)  \\
&  = \int_0^\infty \frac{b_{T}^2 db_{T}}{2\pi} M J_1(b_{T}  k_{T})g_{1T}^\perp(x,b_{T};\mu,\zeta), \end{aligned}
\end{equation}
\begin{equation}
D_1(z,p_{T};\mu,\zeta) = \int_0^\infty \frac{b_{T}db_{T}}{2\pi} J_0(b_{T} p_{T}) D_1(z,b_{T};\mu,\zeta),    
\end{equation}
where $k_{T}$ represents the quark transverse momentum with respect to the nucleon and $p_{T}$ represents the quark transverse momentum with respect to the produced hadron.
The details of the Fourier transformation are given in Appendix \ref{App:Fourier}.

\subsection{Evolution of TMD PDFs and FFs}

The energy scale dependence on $\mu$ and $\zeta$ of the TMD functions are given by the evolution equations,
\begin{gather}
    \label{Equ:Evol mu}
    \mu^2 \frac{d}{d\mu^2} F(x,b_{T};\mu,\zeta) = \frac{\gamma_F(\mu,\zeta)}{2} F(x,b_{T};\mu,\zeta),\\
    \label{Equ:Evol zeta}
    \zeta \frac{d}{d\zeta} F(x,b_{T};\mu,\zeta) = -\mathcal{D}(\mu,b_{T}) F(x,b_{T};\mu,\zeta),
\end{gather}
where $\gamma_F$ is the anomalous dimension, and $\mathcal{D}$ is the rapidity anomalous dimension (RAD), also known as the Collins-Soper kernel.
The $F$ represents some TMD PDF or TMD FF, {\it i.e.} $f_1$, $D_1$, and $g_{1T}^\perp$ in this study. One may have the formal solution,
\begin{equation}
\begin{aligned}
F & (x, b_{T}; \mu, \zeta)  \\
&  =R[\left(b_{T} ; \mu_i, \zeta_i\right)\rightarrow(b_{T} ; \mu, \zeta)] F\left(x, b_{T}; \mu_i, \zeta_i\right),
\end{aligned}
\end{equation}
which relates the TMD PDF (or FF) at $(\mu, \zeta)$ to that at the initial point $(\mu_i,\zeta_i)$. The evolution factor $R\left[\left(b_{T} ; \mu_i, \zeta_i\right) \rightarrow(b_{T} ; \mu, \zeta)\right]$ can be expressed as 
\begin{equation}
\begin{aligned}
&R[(b_{T} ;\mu_i, \zeta_i) \rightarrow(b_{T} ;\mu, \zeta)]\\
& =\exp \left[\int_{\mathcal{P}}\left(\frac{\gamma_F(\mu, \zeta)}{\mu} d \mu -\frac{\mathcal{D}(\mu, b_{T})}{\zeta} d \zeta\right)\right],
\end{aligned}
\end{equation}
where $\mathcal{P}$ represents the path connecting the scales $(\mu_i, \zeta_i)$ and $(\mu, \zeta)$. As a common choice, we set the energy scales as $\mu^2 = \zeta = Q^2$.

According to the integrability condition \cite{Collins:1981va}
\begin{equation}\label{Equ:integrability condition}
\zeta \frac{d}{d \zeta} \gamma_F(\mu, \zeta)=-\mu \frac{d}{d \mu} \mathcal{D}(\mu, b_{T})=-\Gamma_{\text {cusp }}(\mu),
\end{equation}
the evolution factor $R[(b_{T} ;\mu_i, \zeta_i) \rightarrow(b_{T} ;Q, Q^2)]$ is in principle path independent. However, it differs from path to path when truncating at some fixed order in perturbation theory.  As suggested in Ref.~\cite{Scimemi:2019cmh}, the condition~\eqref{Equ:integrability condition} allows one to
construct a two-dimensional field ${\cal F}(\mu,\zeta)$, of which the gradient is given by $\mathbf{E}=(\gamma_F/2,-\mathcal{D})$. Then $F(x,b_T,\mu,\zeta)$ remains unchanged if the path is along the equipotential line of ${\bf E}$, referred to as a null-evolution line. In the $(\mu,\zeta)$ plane, there is a unique saddle point $(\mu_0, \zeta_0)$ defined by
\begin{equation}
    \mathcal{D}(\mu_0, b_{T}) = 0,
    \quad
    \gamma_{F}(\mu_0,\zeta_0) = 0.
\end{equation}
Among the null-evolution lines, only the one passing through the saddle point has finite $\zeta$ at all values of $\mu$. Hence, the $F(x,b_{T}) \equiv F(x,b_{T};\mu_{0},\zeta_{0})$ is referred to as the optimal TMD PDF or FF~\cite{Scimemi:2019cmh}. 
Owing to the good properties of the null-evolution line and the saddle point, we firstly evolve the $F$ from the saddle point along the null-evolution line to the point with $\mu=Q$; secondly, we 
evolve the $F$ along the straight line keeping $\mu=Q$ fixed until reaching the point with $\zeta = Q^2$.
The result for the evolution factor $R[(b_{T} ;\mu_i, \zeta_i) \rightarrow(b_{T} ;Q, Q^2)]$ along this path is \cite{Scimemi:2019cmh}
\begin{equation}
R\left[\left(b_{T} ;\mu_i, \zeta_i\right) \rightarrow\left(b_{T} ;Q, Q^2\right)\right]=\left(\frac{Q^2}{\zeta_\mu(Q, b_{T})}\right)^{-\mathcal{D}(Q, b_{T})}.
\end{equation}
The expressions for $\mathcal{D}(Q, b_{T})$ and $\zeta_\mu(Q, b_{T})$ can be found in Appendix \ref{App:Evolution}.
The precision for the perturbative calculation of various factors in powers of $\alpha_s$ in this work is summarized in Table \ref{tab:alpha order}.

\begin{table}
	\caption{Orders of perturbative calculations for anomalous dimensions and the $C(\mathbb{C})$ functions in the optimal TMD PDF and FF.}
 \label{tab:alpha order}
\begin{tabular}{c|cccccc}
\hline\hline
        & $\Gamma_{\rm cusp}$ & $\gamma_V$  & $\mathcal{D}_{\rm resum}$ & $\zeta_{\mu}^{\rm pert}$ & $\zeta_{\mu}^{\rm exact}$    \\
$R$     &   $\alpha_s^3$  & $\alpha_s^2$ & $\alpha_s^2$ & $\alpha_s^1$ & $\alpha_s^1$ \\ \hline\hline
& $f_1$ & $D_1$ & $g_{1T}^{\perp}$ \\
$C(\mathbb{C})$ & $\alpha_s^1$ & $\alpha_s^1$ & $\alpha_s^0$
\\ \hline\hline
\end{tabular}
\end{table}

\subsection{Unpolarized TMD PDF and FF} \label{subsec: OPT Unpol PDF FF}

For unpolarized TMD PDFs and FFs, we adopt the SV19 parametrization~\cite{Scimemi:2019cmh}. 
The optimal unpolarized TMD PDF and FF are expressed as
\begin{align}
    &f_{1, f \leftarrow h}\left(x, b_{T}\right)= \sum_{f^{\prime}} \int_x^1 \frac{d y}{y} C_{f \leftarrow f^{\prime}}\left(y, b_{T}, \mu_{\mathrm{OPE}}^{\mathrm{PDF}}\right)
    \notag\\
    &\quad\quad \times f_{1, f^{\prime} \leftarrow h}\left(\frac{x}{y}, \mu_{\mathrm{OPE}}^{\mathrm{PDF}}\right) f_{\mathrm{NP}}(x, b_{T}), \\
    &D_{1, f \rightarrow h}\left(z, b_{T}\right)= \frac{1}{z^2} \sum_{f^{\prime}} \int_z \frac{d y}{y} y^2 \mathbb{C}_{f \rightarrow f^{\prime}}\left(y, b_{T}, \mu_{\mathrm{OPE}}^{\mathrm{FF}}\right)
    \notag\\
    &\quad\quad\times d_{1, f^{\prime} \rightarrow h}\left(\frac{z}{y}, \mu_{\mathrm{OPE}}^{\mathrm{FF}}\right) D_{\mathrm{NP}}(z, b_{T}),
\end{align}
where $f_{1,f^\prime \leftarrow h}$ and $d_{1,f^\prime \rightarrow h}$ are collinear PDFs and FFs. The scales $\mu_{\mathrm{OPE}}^{\mathrm{PDE}}$ and $\mu_{\mathrm{OPE}}^{\mathrm{FF}}$ are chosen as
\begin{gather}
    \mu_{\mathrm{OPE}}^{\mathrm{PDE}}=\frac{2 e^{-\gamma_E}}{b_{T}}+2\,\mathrm{GeV},\\
    \mu_{\mathrm{OPE}}^{\mathrm{FF}}=\frac{2 e^{-\gamma_E} z}{b_{T}}+2\,\mathrm{GeV},
\end{gather}
where $\gamma_E$ is the Euler-Mascheroni constant. The $2\,\mathrm{GeV}$ shift is introduced to keep the PDFs and FFs in perturbative region when $b_T$ is large. The nonperturbative functions $f_{\rm NP}(x, b_T)$ and $D_{\rm NP}(z,b_T)$ are to be parametrized.

For unpolarized TMD PDF, the coefficient function $C$ can be written as
\begin{equation}
 \begin{aligned}
    C_{f \leftarrow f^{\prime}}(x, b_{T}, \mu)= & \delta(1-x) \delta_{f f^{\prime}}+ \\ 
    &  a_s(\mu)\left(-\mathbf{L}_\mu P_{f \leftarrow f^{\prime}}^{(1)} +C_{f \leftarrow f^{\prime}}^{(1,0)}\right),
 \end{aligned}
\end{equation}
up to NLO, where $a_s = \frac{g^2(\mu)}{(4\pi)^2}$ and $g(\mu)$ are the QCD coupling constants. $\mathbf{L}_\mu$ is defined as
\begin{equation}
\mathbf{L}_\mu=\ln \left(\frac{b_{T}^2 \mu^2}{4 e ^{-2 \gamma_E}}\right).
\end{equation}
$P_{f \leftarrow f^{\prime}}^{(1)}$ is the coefficient of the PDF evolution kernel, which reads
\begin{gather}
P_{q \leftarrow q^{\prime}}^{(1)}(x)=2 C_F\left(\frac{1+x^2}{1-x}\right)_+ \delta_{q q^{\prime}}, \\
P_{q \leftarrow g}^{(1)}(x)=1-2 x+2 x^2 .
\end{gather}
The ``$+$'' prescription is defined as
\begin{equation}
\int_{x_0}^1 d x[g(x)]_{+} f(x)=\int_0^1 d x g(x)\left[f(x) \Theta\left(x-x_0\right)-f(1)\right],
\end{equation}
where $\Theta\left(x-x_0\right)$ is the Heaviside step function.
$C_F = 4/3$ is the quadratic Casimir eigenvalues of fundamental representation of $SU(3)$. 
The expressions of $C_{f \leftarrow f^{\prime}}^{(n,0)}$ can be found in Ref. \cite{Zeng:2022lbo}, and their NLO terms are
\begin{align}
C_{q \leftarrow q^{\prime}}^{(1,0)}(x)&=C_F\left(2 \bar{x}-\delta(\bar{x}) \frac{\pi^2}{6}\right)\delta_{qq^{\prime}}, \\
C_{q \leftarrow g}^{(1,0)}(x) &=2 x \bar{x} ,
\end{align}
and $\bar{x} = 1-x$.

For unpolarized TMD FF, one should replace $C_{f \leftarrow f^{\prime}}(x, b_{T}, \mu)$, $P_{f \leftarrow f^{\prime}}^{(1)}$ and $C_{f \leftarrow f^{\prime}}^{(n,0)}(x)$ by $\mathbb{C}_{f \rightarrow f^{\prime}}\left(z, b_{T}, \mu\right)$, $\mathbb{P}_{f \rightarrow f^{\prime}}^{(1)}$ and $\mathbb{C}_{f \rightarrow f^{\prime}}^{(n,0)}(z)$, which at NLO are expressed as
\begin{align}
\mathbb{P}_{q \rightarrow q^{\prime}}^{(1)}(z)&=\frac{2 C_F}{z^2}\bigg(\frac{1+z^2}{1-z}\bigg)_{+} \delta_{q q^{\prime}}, \\
\mathbb{P}_{q \rightarrow g}^{(1)}(z)&=\frac{2 C_F}{z^2} \frac{1+(1-z)^2}{z}, \\
\mathbb{C}_{q \rightarrow q^{\prime}}^{(1,0)}(z)&=\frac{C_F}{z^2}  \bigg[2(1-z) + 
  \frac{4(1+z^2) \ln z}{1-z} \\
 &\quad
 -\delta(1-z) \frac{\pi^2}{6}\bigg] \delta_{q q^{\prime}},\\
\mathbb{C}_{q \rightarrow g}^{(1,0)}(z)&=\frac{2 C_F}{z^2}\bigg[z+2\big(1+(1-z)^2\big) \frac{\ln z}{z}\bigg].
\end{align}

\begin{widetext} 
\subsection{Worm-gear asymmetry}

The longitudinal-transverse double spin asymmetry of the SIDIS process is defined as
\begin{align}
A_{L T}= & \frac{1}{\left|S_{\perp}\right|\left|\lambda_e\right|} \frac{\left[\mathrm{d} \sigma_{L T}(+, \uparrow)-\mathrm{d} \sigma_{L T}(-, \uparrow)\right]-\left[\mathrm{d} \sigma_{L T}(+, \downarrow)-\mathrm{d} \sigma_{L T}(-, \downarrow)\right]}{\mathrm{d} \sigma_{L T}(+, \uparrow)+\mathrm{d} \sigma_{L T}(-, \uparrow)+\mathrm{d} \sigma_{L T}(+, \downarrow)+\mathrm{d} \sigma_{L T}(-, \downarrow)},
\end{align}
where $+$ ($-$) represents the positive (negative) helicity state of the electron beam and $\uparrow$ ($\downarrow$) represents the transverse spin direction of nucleon $S_{\bot}$ to be parallel (anti-parallel) to the designated positive transverse axis. The worm-gear asymmetry is defined as $\cos \left(\phi_h-\phi_S\right)$ modulation of the double spin asymmetry,
\begin{equation}
A_{L T}^{\cos \left(\phi_h-\phi_S\right)}=\frac{\left\langle 2 \cos \left(\phi_h-\phi_S\right) \sigma_{\mathrm{LT}}\right\rangle}{\sqrt{1-\varepsilon^2}\left\langle\sigma_{U U}\right\rangle}=\frac{F_{L T}^{\cos \left(\phi_h-\phi_S\right)}}{F_{U U}} .
\end{equation}
and one can express it with functions we defined above as
\begin{equation}
A_{L T}^{\cos \left(\phi_h-\phi_S\right)}=\frac{M \sum_q e_q^2 \int_0^{\infty} \frac{b_{T}^2 d b_{T}}{2 \pi} J_1\left(\frac{b_{T} P_{hT}}{z}\right) R^2\left[\left(b_{T} ;\mu_i, \zeta_i\right) \rightarrow\left(b_{T} ;Q, Q^2\right)\right] g_{1 T, q \leftarrow N}^\perp(x, b_{T}) D_{1, q \rightarrow h}(z, b_{T})}{\sum_q e_q^2 \int_0^{\infty} \frac{b_{T} d b_{T}}{2 \pi} J_0\left(\frac{b_{T} P_{hT}}{z}\right) R^2\left[\left(b_{T} ;\mu_i, \zeta_i\right) \rightarrow\left(b_{T} ;Q, Q^2\right)\right] f_{1, q \leftarrow N}(x, b_{T}) D_{1, q \rightarrow h}(z, b_{T})} ,
\end{equation}
where $N$ is the target and $h$ is the detected hadron.
\end{widetext}

\section{Extraction of the worm-gear distributions} 
\label{Sec III}

\subsection{Fit to world SIDIS data}

With the formalism above, we perform a global analysis of world SIDIS data to extract the worm-gear distributions $g_{1T}^\perp$ of the nucleon. The results will also serve as the baseline for the impact study of the EicC. 

We parametrize the optimal worm-gear distributions of the proton at the initial scale as
\begin{align}
\label{equ:Fit-B parameterization u d}
g_{1 T,q\leftarrow p}^\perp (x, b_{T}) =N_q \frac{(1-x)^{\alpha_q} x^{\beta_q}}{\mathrm{~B}(\alpha+1, \beta+1)} \exp \left(-r_q b_{T}^2\right)
\end{align}
for $u$ and $d$ quarks and
\begin{align}
\label{equ:Fit-B parameterization sea}
g_{1 T,q\leftarrow p}^\perp (x, b_{T}) =N_q f_1(x, \mu_0) \exp \left(-r_q b_{T}^2\right)
\end{align}
for $\bar u$, $\bar d$, $s$, and $\bar s$ quarks with $\mu_0=2\,\rm GeV$. Here $\mathrm{~B}(\alpha+1, \beta+1)$ is the Euler Beta function, introduced to reduce the correlation among parameters. Assuming the isospin symmetry, we can express corresponding distribution functions of the neutron as
\begin{align}
g_{1 T,u \leftarrow n}^\perp(x, b_{T}) & =g_{1 T,d \leftarrow P}^\perp (x, b_{T}),\\
g_{1 T,\bar{u}  \leftarrow n}^\perp (x, b_{T}) & = g_{1 T,,\bar{d} \leftarrow P}^\perp (x, b_{T}), \\
g_{1 T,d \leftarrow n}^\perp(x, b_{T}) & =g_{1 T,u \leftarrow P}^\perp (x, b_{T}), \\
g_{1 T,\bar{d}  \leftarrow n}^\perp (x, b_{T}) & = g_{1 T,\bar{u} \leftarrow P}^\perp (x, b_{T}), \\
g_{1 T,s \leftarrow n}^\perp(x, b_{T}) & =g_{1 T,s \leftarrow P}^\perp (x, b_{T}), \\
g_{1 T,\bar{s}  \leftarrow n}^\perp (x, b_{T}) & = g_{1 T,\bar{s} \leftarrow P}^\perp (x, b_{T}).
\end{align}

For unpolarized TMD PDFs and FFs, we adopt the SV19 fit~\cite{Scimemi:2019cmh}, in which the nonperturbative functions $f_{\mathrm{NP}}$ and $D_{\mathrm{NP}}$ are parametrized as
\begin{gather}
 f_{\mathrm{NP}}(x, b_{T})=\exp \left[-\frac{\lambda_1(1-x)+\lambda_2 x+x(1-x) \lambda_5}{\sqrt{1+\lambda_3 x^{\lambda_4} b_{T}^2}} b_{T}^2\right], \\
 D_{\mathrm{NP}}(z, b_{T})=\exp \left[-\frac{\eta_1 z+\eta_2(1-z)}{\sqrt{1+\eta_3(b_{T} / z)^2}} \frac{b_{T}^2}{z^2}\right]\left(1+\eta_4 \frac{b_{T}^2}{z^2}\right).
\end{gather}
The values of the parameters $\lambda_i$ and $\eta_i$ are listed in Table~\ref{tab:SV19 para},
which can be also found in \cite{Scimemi:2019cmh,Zeng:2022lbo}.
For the FFs to charged hadrons, we approximate them as 
\begin{align}
D_{1,f\leftarrow h^{+}} & = D_{1,f\leftarrow \pi^{+}} + D_{1,f\leftarrow K^{+}} + D_{1,f\leftarrow p},\\
D_{1,f\leftarrow h^{-}} & = D_{1,f\leftarrow \pi^{-}} + D_{1,f\leftarrow K^{-}} + D_{1,f\leftarrow \bar{p}}.
\end{align} 

\begin{table}
	\caption{The parameters for nonperturbative functions of the optimal unpolarized TMD PDF and FF. The units are in $\mathrm{GeV}^2$ except that $\lambda_4$ is dimensionless. 
 }
        \label{tab:SV19 para}
\begin{tabular}{ccccccccc}
\hline\hline
$\lambda_1$ & $\quad$ & $\lambda_2$ & $\quad$ & $\lambda_3$ & $\quad$ & $\lambda_4$ & $\quad$ & $\lambda_5$ \\
$ 0.198 $   & $\quad$ & $ 9.3$      &  $\quad$ &  $ 431$     &  $\quad$ & $ 2.12 $ & $\quad$ & $ -4.44$    \\\hline\hline
$\eta_1$    & $\quad$ & $\eta_2$    & $\quad$ & $\eta_3$    & $\quad$ & $\eta_4$     \\
$ 0.260$   & $\quad$ & $ 0.476$      & $\quad$ & $ 0.478$     & $\quad$ & $ 0.483$       \\\hline\hline
\end{tabular}

\end{table}

\begin{table*}
	\caption{The SIDIS double spin asymmetry data by HERMES~\cite{HERMES:2020ifk}, COMPASS~\cite{COMPASS:2016led,Parsamyan:2018evv,Avakian:2019drf}, and JLab~\cite{JeffersonLabHallA:2011vwy}. The SFA refers to $\left\langle 2 \cos \left(\phi_h-\phi_S\right) \sigma_{LT}\right\rangle/\left(\sqrt{1-\varepsilon^2}\left\langle\sigma_{U U}\right\rangle\right)$, and the CSA refers to $\left\langle 2 \cos \left(\phi_h-\phi_S\right) \sigma_{LT}\right\rangle/\left\langle\sigma_{U U}\right\rangle$.}
        \label{tab:fit-A data}
	\begin{ruledtabular}
\begin{tabular}{llcllllcll}
Data set        & Target    & Beam               & original data &  data points  &  data points   & Process     & Measurement  \\
               &            &                   & points  &  after cut  &  after cut \\
             &            &                   &          & $\delta < 0.5,$ & $\delta < 0.3,$ \\
             &            &                   &          & $Q>1\,\mathrm{GeV}$ & $Q>1\,\mathrm{GeV}$ \\
\\\hline
HERMES \cite{HERMES:2020ifk}          & H$_2$     & $27.6\,\mathrm{GeV}\, e^{\pm}$  &          64  & 26 & 11 & $e^{\pm}p\rightarrow e^{\pm} \pi^{+}X$ & SFA\\
                &           &                    &          64  & 26 & 11 & $e^{\pm}p\rightarrow e^{\pm} \pi^{-}X$ & SFA\\
                &           &                    &          64  & 26 & 12 & $e^{\pm}p\rightarrow e^{\pm} K^{+}X$ & SFA\\
                &           &                    &          64  & 26 & 12 & $e^{\pm}p\rightarrow e^{\pm} K^{-}X$ & SFA\\
                &           &                    &          64  & 30 & 15  & $e^{\pm}p\rightarrow e^{\pm} PX$ & SFA\\ \hline
COMPASS \cite{COMPASS:2016led}        & NH$_3$    & $160\,\mathrm{GeV}\,\mu^{+}$   &        66  & 28 & 9  & $\mu^{+}p\rightarrow \mu^{+} h^+X$ & SFA \\
                &           &                    &          66  & 26 & 8  & $\mu^{+}p\rightarrow \mu^{+} h^-X$ & SFA \\ \hline
JLab  \cite{JeffersonLabHallA:2011vwy}          & $^3$He    & $5.9\,\mathrm{GeV}\, e^{-}$     &          4   & 2 & 1     &   $e^{-}n\rightarrow e^{-} \pi^{+}X$& CSA\\
                &           &                    &          4   & 2 & 1     &   $e^{-}n\rightarrow e^{-} \pi^{-}X$& CSA\\\hline
Total           &           &                    &          460 & 192 & 80
&\\    
\end{tabular}
\end{ruledtabular}
\end{table*}

In this analysis, we include the SIDIS longitudinal-transverse double spin asymmetry data from HERMES~\cite{HERMES:2020ifk}, COMPASS~\cite{COMPASS:2016led,Parsamyan:2018evv,Avakian:2019drf} and JLab~\cite{JeffersonLabHallA:2011vwy}, as summarized in Table~\ref{tab:fit-A data}. Since the TMD factorization is only valid at small $\delta = |P_{hT}|/(zQ)$, only data with $\delta < 0.5$ are included in the fit. 

For the HERMES data, experimental results are provided in both one-dimensional binning and three-dimensional binning. We only use the three-dimensional bins in this study, because they are supposed to contain more information for the study of TMDs, which are multidimensional functions. For the COMPASS data, the experimental results are provided in one-dimensional binning but on $x$, $z$, and $P_{hT}$ respectively. Since they tell the dependence on different variable, we include all these bins in the fit. However, to avoid double counting, we multiply a factor of $1/3$ when calculating the $\chi^2$ from the COMPASS data. Then the total $\chi^2/N$ to be minimized in the fit is defined as
\begin{equation}
    \label{eq::total chi2 Fit A }
    \chi^2/N = \frac{\frac{1}{3}\chi^2_{\mathrm{COMPASS}} + \chi^2_{\mathrm{HERMES}} + \chi^2_{\mathrm{JLab}}}{\frac{1}{3}N_{\mathrm{COMPASS}} + N_{\mathrm{HERMES}} + N_{\mathrm{JLab}}},
\end{equation}
where $N_{\rm data~set}$ represents the number of points for each data set. For each data set, we have
\begin{equation}
\chi^2_{\text {data set}}= \sum_{\substack{i, j \in\\ \text {data-points }}}\left(t_i-a_i\right) V_{i j}^{-1}\left(t_j-a_j\right),
\end{equation}
where $i$ and $j$ run over all points in each set, $t_i$ represent the theoretical values, and $a_i$ represent experimental values. The $V$-matrix is given by
\begin{equation}
V_{i j}=\delta_{i j} \left(\sigma_i^{\rm uncor.}\right)^2 + \sigma_i^{\rm cor.} \sigma_j^{\rm cor.},
\end{equation}
where $\sigma_i^{\text { uncor. }}$ and $\sigma_i^{\rm cor.}$ stand for uncorrelated and correlated uncertainties respectively. 

As the existing world data are not precise enough to constrain all parameters introduced in Eq. (\ref{equ:Fit-B parameterization u d}), we practically reduce the number of parameters by imposing the conditions, 
\begin{equation}
\alpha_{u}=\alpha_{d}=\alpha, \quad \beta_{u}=\beta_{d}=\beta, \quad r_{u}=r_{d}=r,
\end{equation}
and assuming vanishing distributions for sea quarks, $\bar{u}$, $\bar{d}$, $s$, and $\bar{s}$. 
In the end, we have five free parameters to be determined as listed in Table~\ref{tab:fit-A parameters}.

To estimate the uncertainties, we create 1000 replicas of the data by smearing the central values of each data point according to a Gaussian distribution with data uncertainties being the widths. For each replica, we perform a fit. Then the central values of all physical quantities are evaluated from the average of the 1000 fits. More details of this approach are described in Ref.~\cite{Zeng:2022lbo}.

\begin{figure*}[htbp]
    \centering
    \includegraphics[scale=0.7]{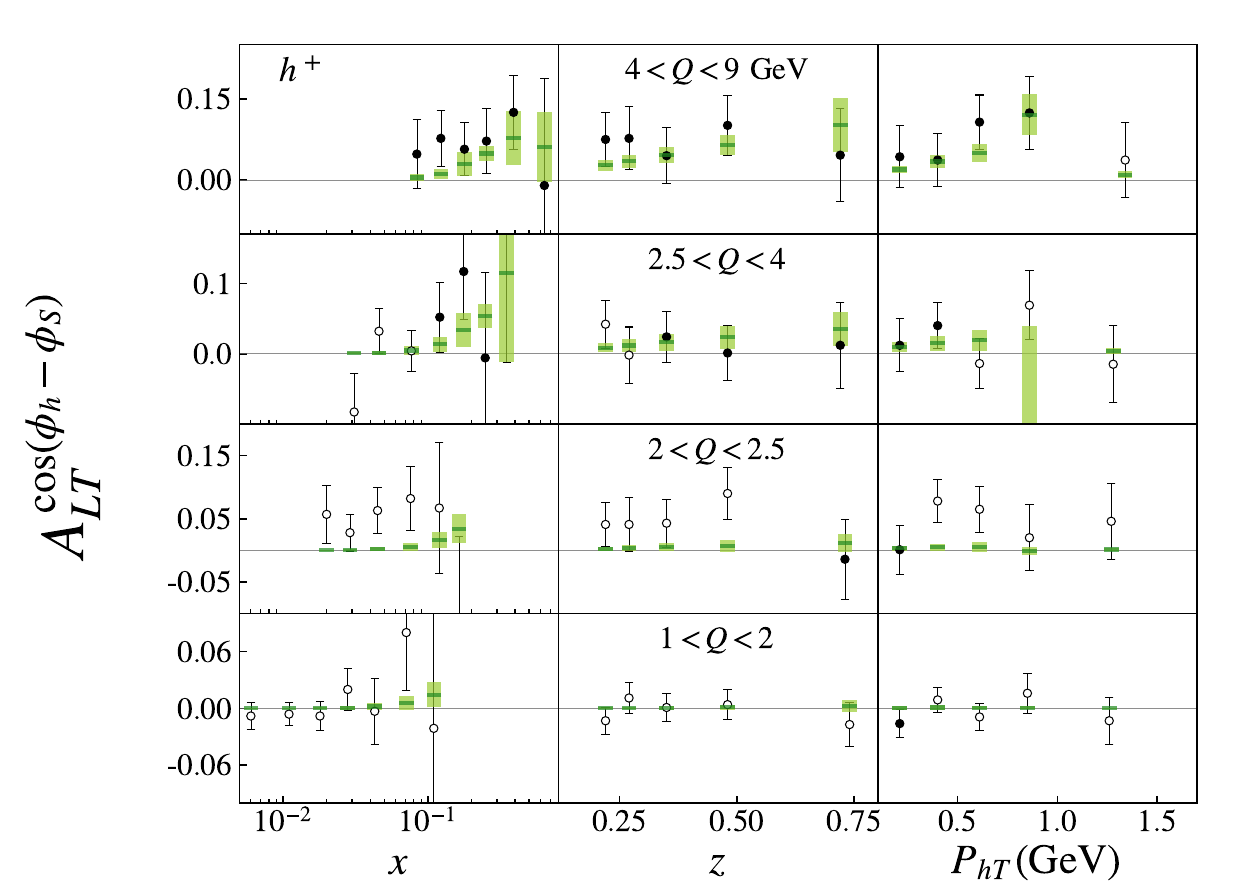}
    \includegraphics[scale=0.7]{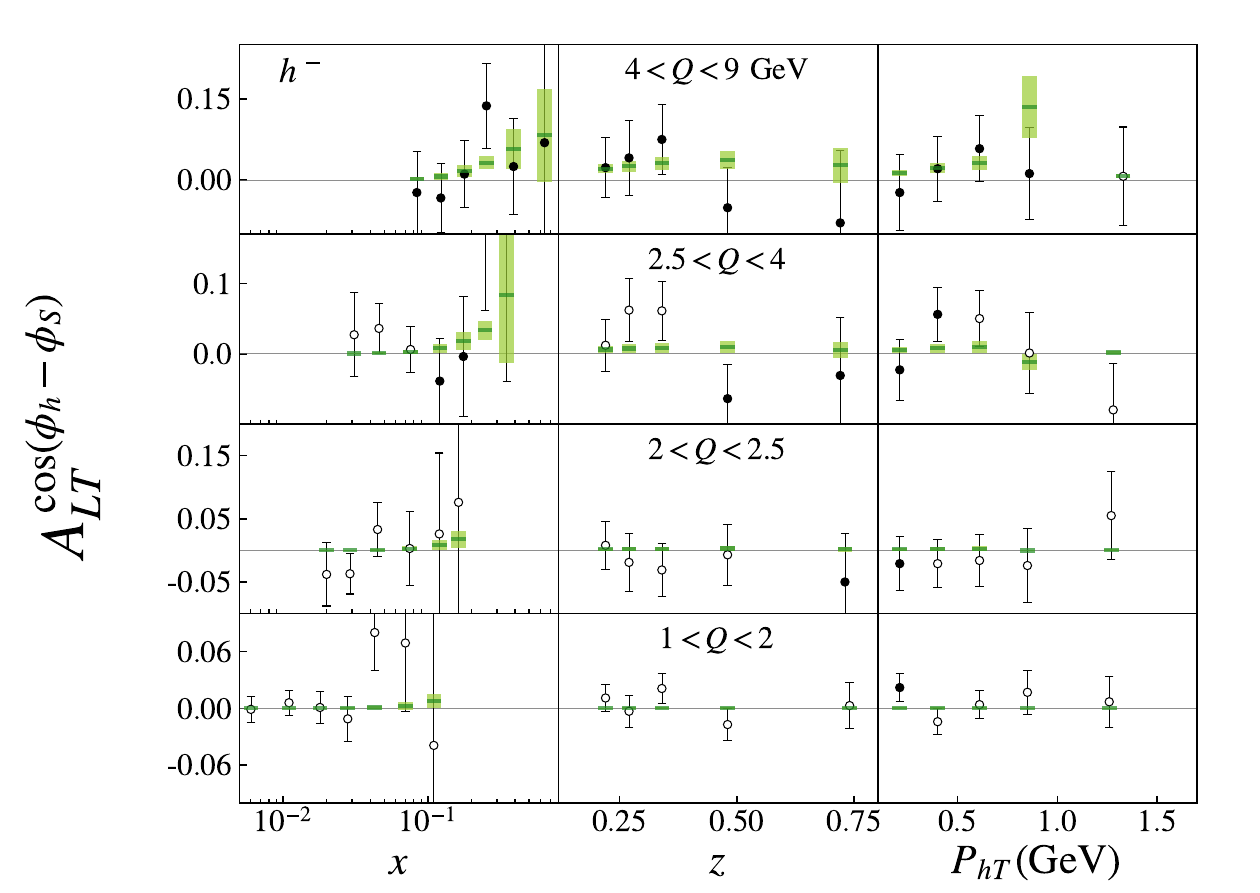}
    \caption{Comparison between the fit results and the experimental data by COMPASS~\cite{COMPASS:2016led} from the proton target with charged hadron $h^{\pm}$ measured in the final state. The filled points are within the kinematic cuts, $\delta<0.5$ and $Q>1\,\rm GeV$, and included the fit, while the open points are not included in the fit. The green lines and the bands are the mean values and the standard deviations calculated from the fits to 1000 replicas.}
    \label{Fig:COMPASS hm}
    \end{figure*}

\begin{figure*}[htbp]
    \centering
    \includegraphics[scale=0.7]{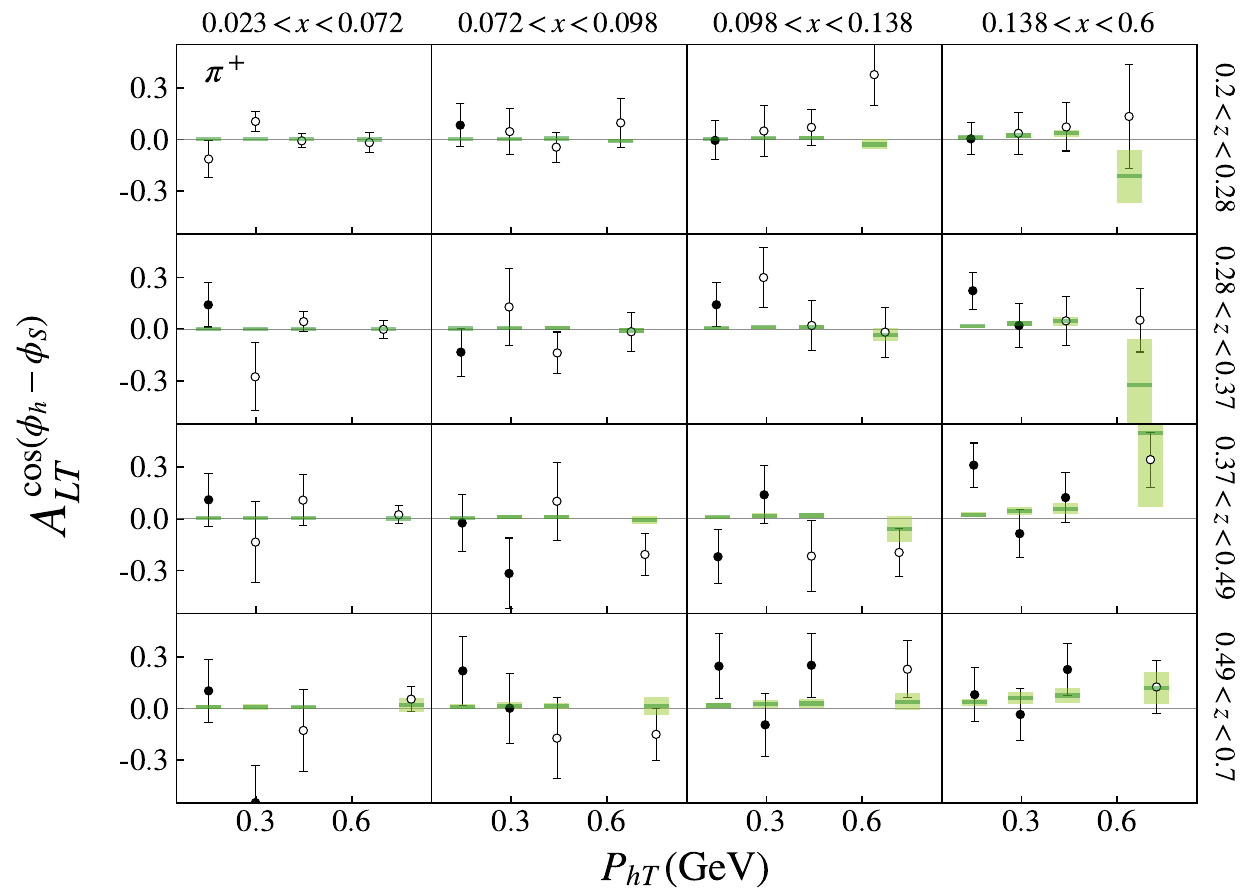}
    \includegraphics[scale=0.7]{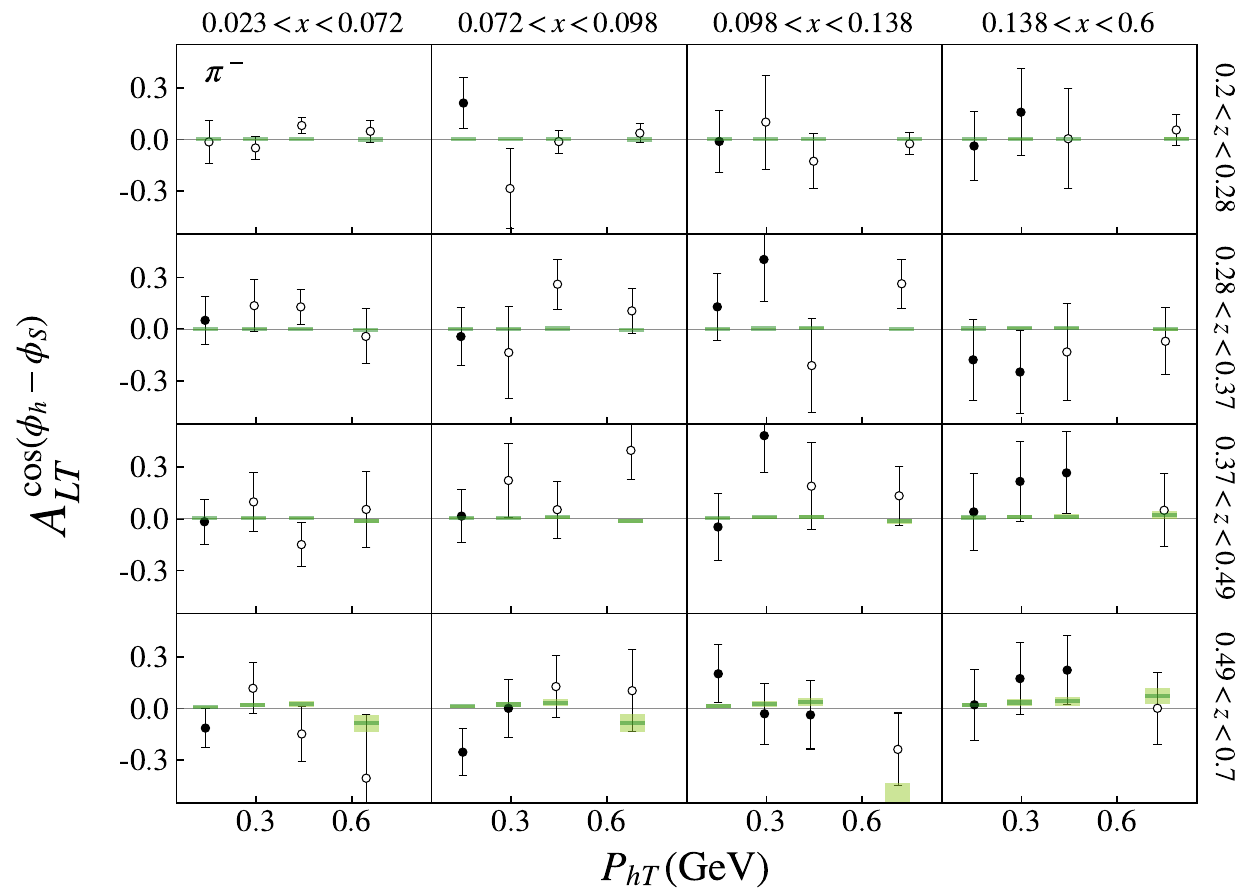}
    \caption{Comparison between the fit results and the experimental data by HERMES \cite{HERMES:2020ifk} from the proton target with $\pi^{\pm}$ measured in the final state. The data points and the fit bands follow the same notations as Fig.~\ref{Fig:COMPASS hm}.}
    \label{Fig:HERMES pin}
    \end{figure*}

\begin{figure*}[htbp]
    \centering
    \includegraphics[scale=0.7]{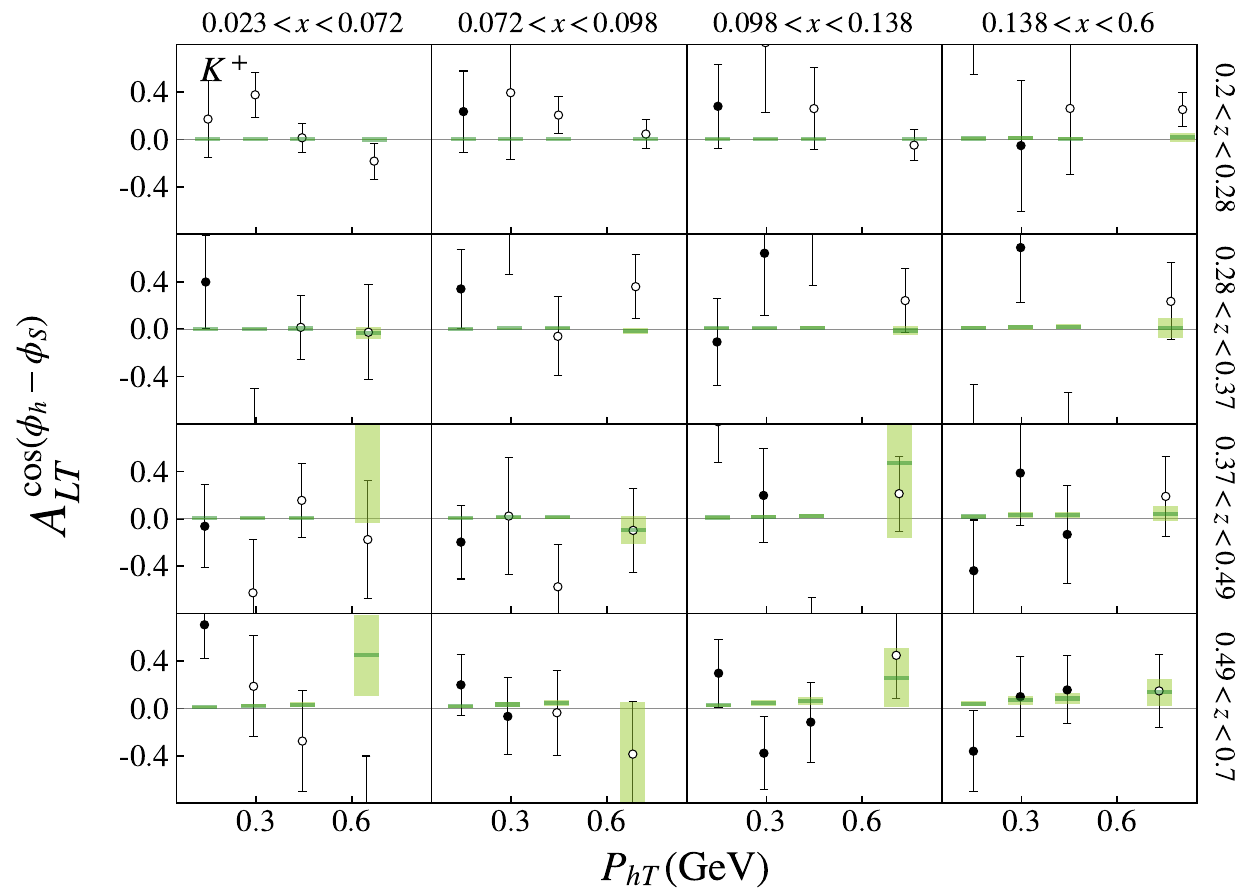}
    \includegraphics[scale=0.7]{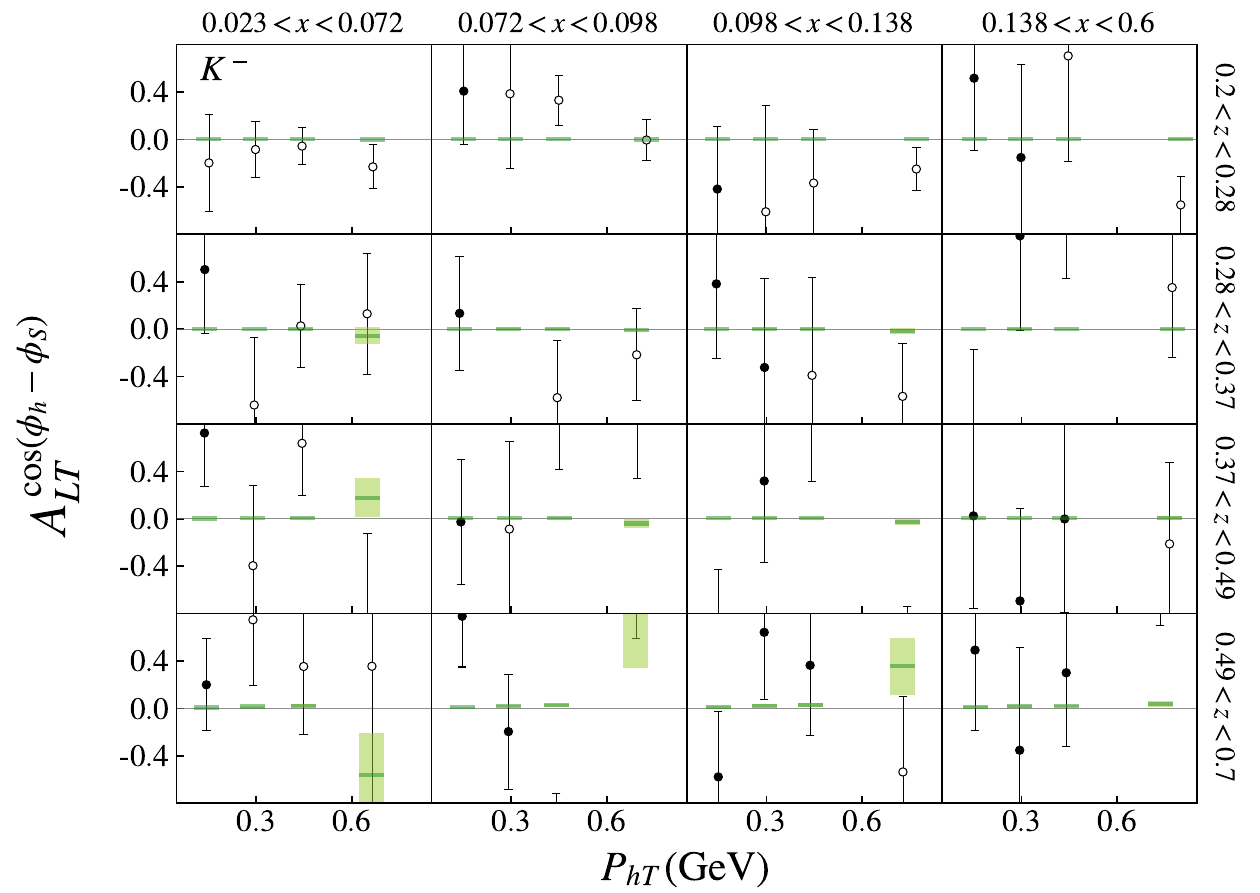}
    \caption{Comparison between the fit results and experimental data by HERMES~\cite{HERMES:2020ifk} from the proton target with $K^{\pm}$ measured in the final state.
    The data points and the fit bands follow the same notations as Fig.~\ref{Fig:COMPASS hm}.}
    \label{Fig:HERMES Kn}
    \end{figure*}

\begin{figure*}[htbp]
    \centering
    \includegraphics[scale=0.7]{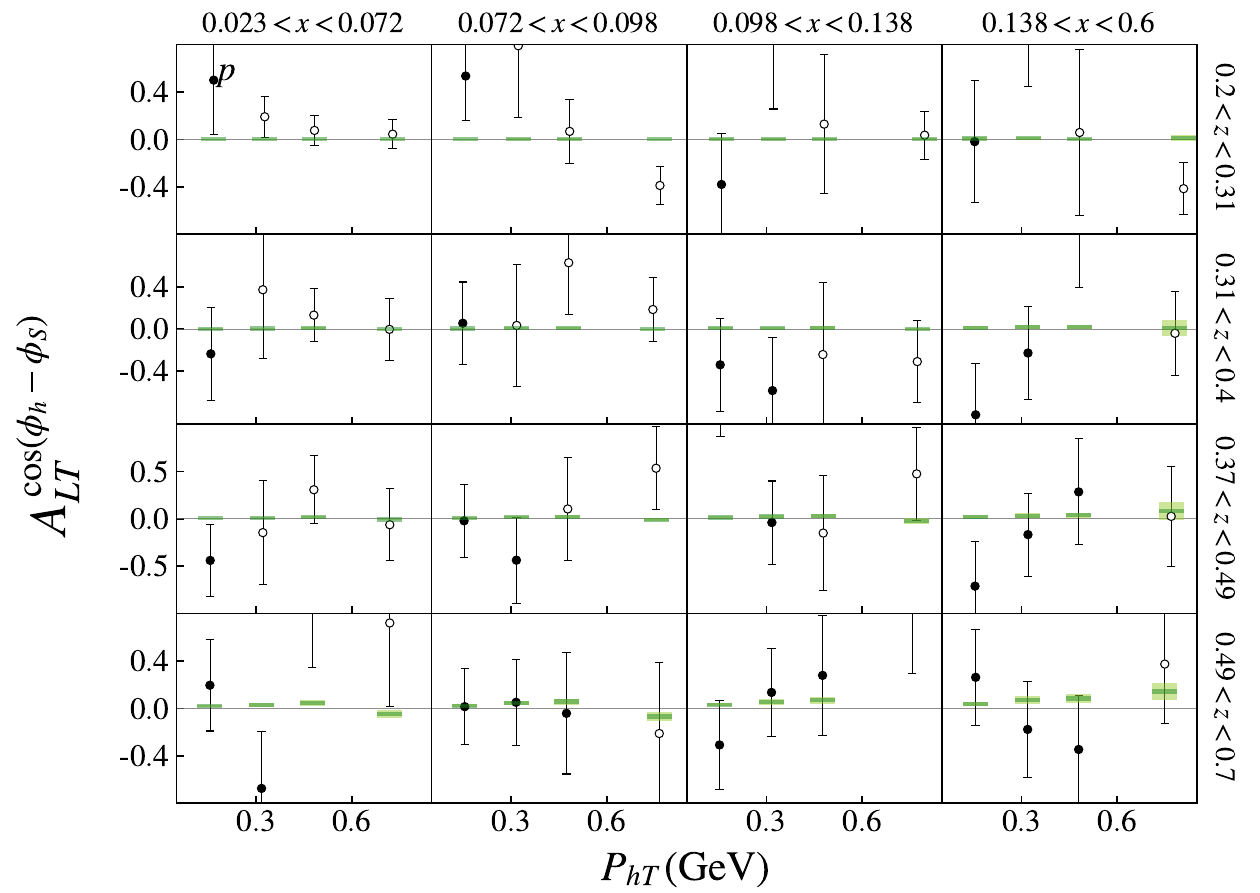}
        \caption{Comparison between the fit results and the experimental data by HERMES~\cite{HERMES:2020ifk} from the proton target with a fragmented proton measured in the final state. The data points and the fit bands follow the same notation as Fig.~\ref{Fig:COMPASS hm}.}
    \label{Fig:HERMES P}
    \end{figure*}

\begin{figure*}[htbp]
    \centering
    \includegraphics[scale=0.5]{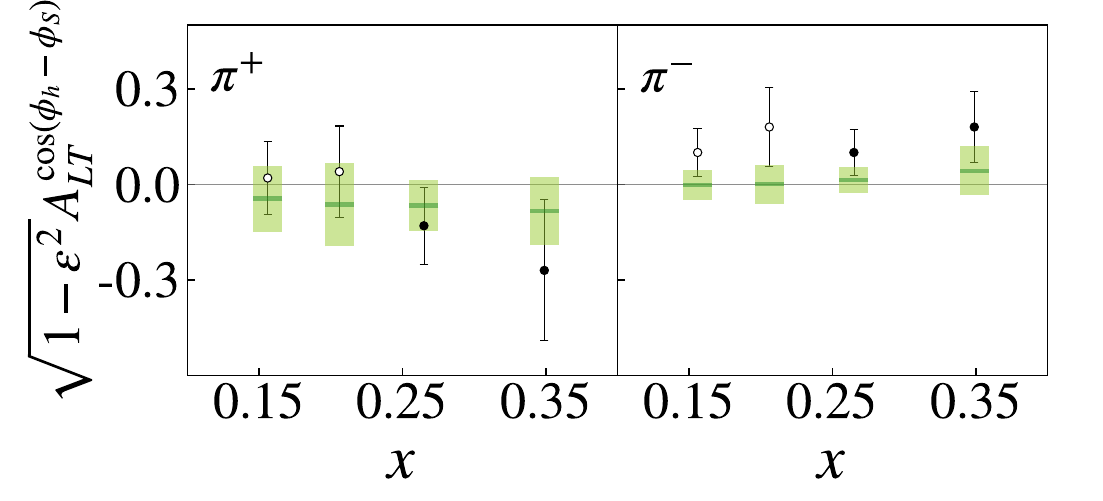}
    \caption{Comparison between the fit results and the experimental data by JLab~\cite{JeffersonLabHallA:2011vwy} from the effectively polarized neutron target with $\pi^{\pm}$ measured in the final state. The data points and the fit bands follow the same notation as Fig.~\ref{Fig:COMPASS hm}.}
    \label{Fig:JLab}
    \end{figure*}

\begin{figure*}[htbp]
  \centering
  {\includegraphics[width=0.32\textwidth,page = 1]{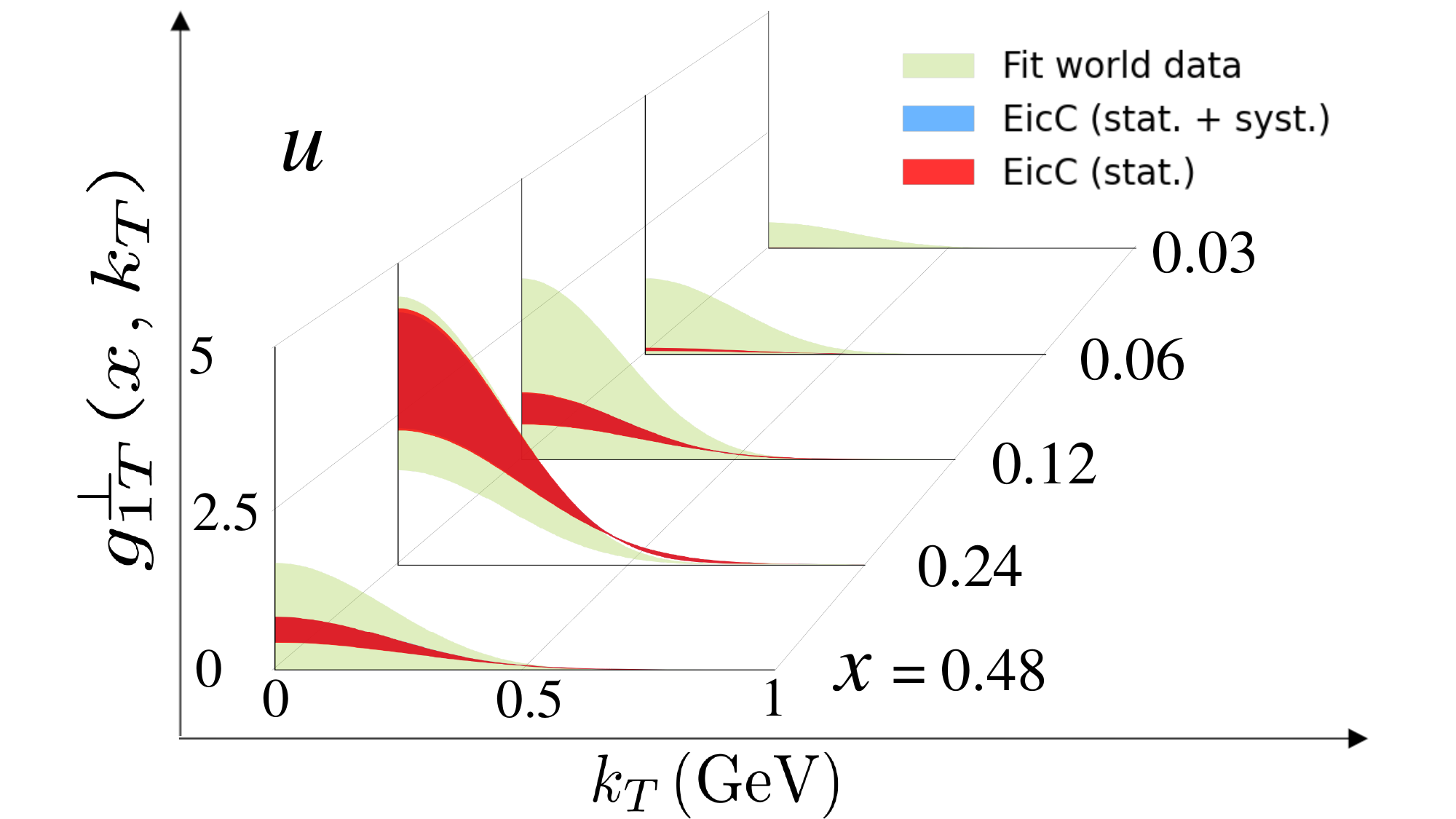}}
  {\includegraphics[width=0.32\textwidth,page = 2]{figures/g1T-x-kt.pdf}}
  {\includegraphics[width=0.32\textwidth,page = 3]{figures/g1T-x-kt.pdf}}
  {\includegraphics[width=0.32\textwidth,page = 4]{figures/g1T-x-kt.pdf}}
  {\includegraphics[width=0.32\textwidth,page = 5]{figures/g1T-x-kt.pdf}}
  {\includegraphics[width=0.32\textwidth,page = 6]{figures/g1T-x-kt.pdf}}
  \caption{The worm-gear distributions $g_{1T}^\perp(x,k_T)$ at the scale $Q=2\,\rm GeV$. The uncertainty bands correspond to 68\% CL estimated from the fits to 1000 replicas. The green bands are extracted distributions by fitting the world SIDIS data, the red bands are EicC projections with only statistical uncertainties, and the blue bands are EicC projections with both statistical and systematic uncertainties.}
    \label{Fig:xg1Txkt}
\end{figure*}

\begin{figure*}[htbp]
  \centering
  {\includegraphics[scale=0.6]{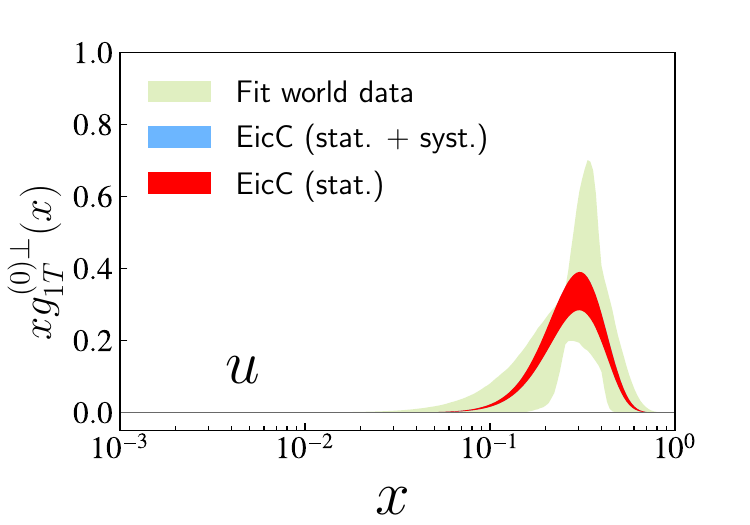}}
  {\includegraphics[scale=0.6]{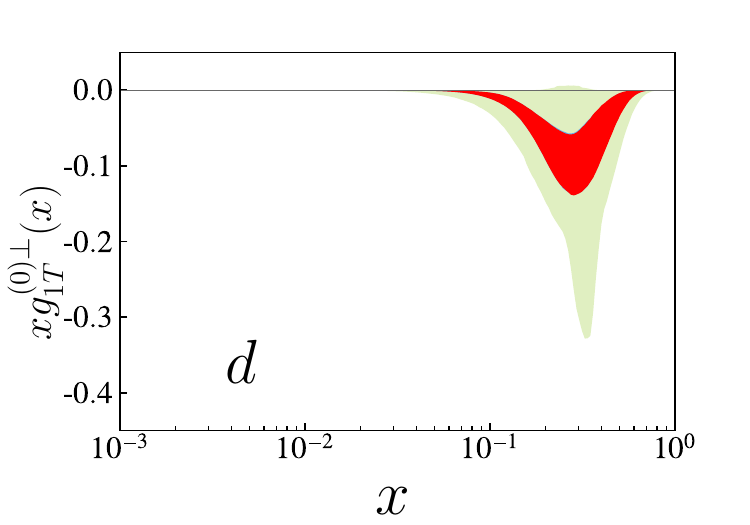}}
  \caption{The zeroth transverse moment of the worm-gear functions, $g_{1T}^{\perp(0)}(x)$ as defined in Eq.~\eqref{equ::g1T0x}, for $u$ and $d$ quarks at the scale $Q=2\,\rm GeV$. The uncertainty bands correspond to 68\% CL estimated from the fits to 1000 replicas. The green bands are extracted distributions by fitting the world SIDIS data, the red bands are EicC projections with only statistical uncertainties, and the blue bands are EicC projections with both statistical and systematic uncertainties.}
    \label{Fig:xg1Tx}
\end{figure*}

\begin{figure*}[htbp]
  \centering
  {\includegraphics[scale=0.55]{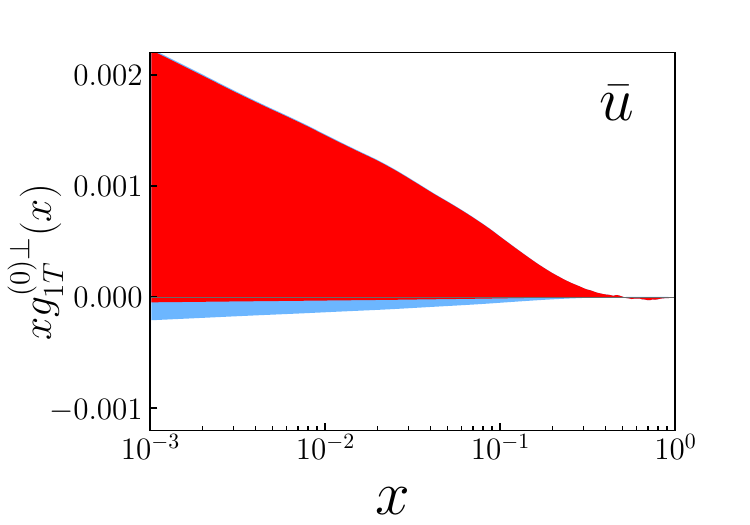}}
  {\includegraphics[scale=0.55]{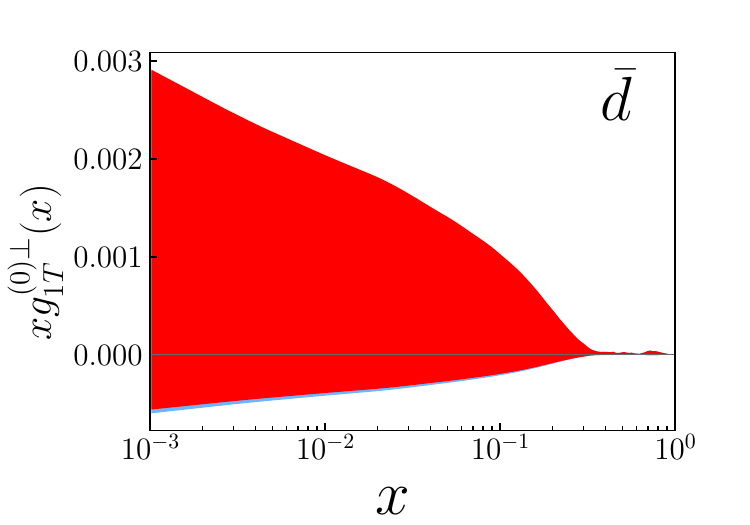}}
  {\includegraphics[scale=0.55]{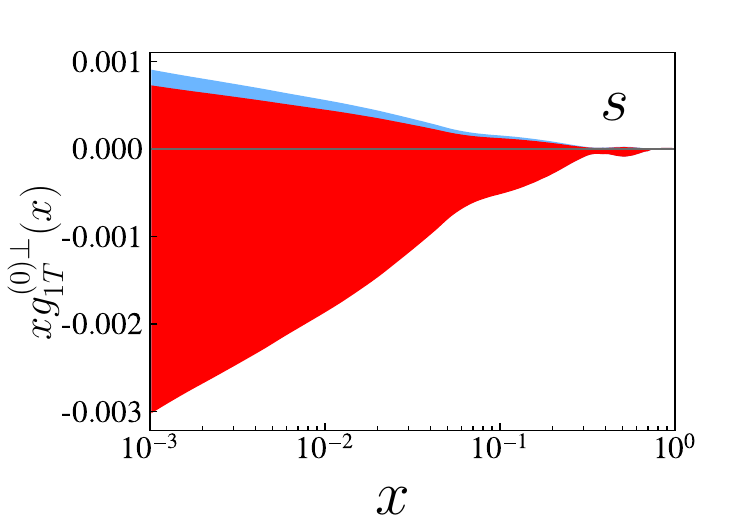}}
  {\includegraphics[scale=0.55]{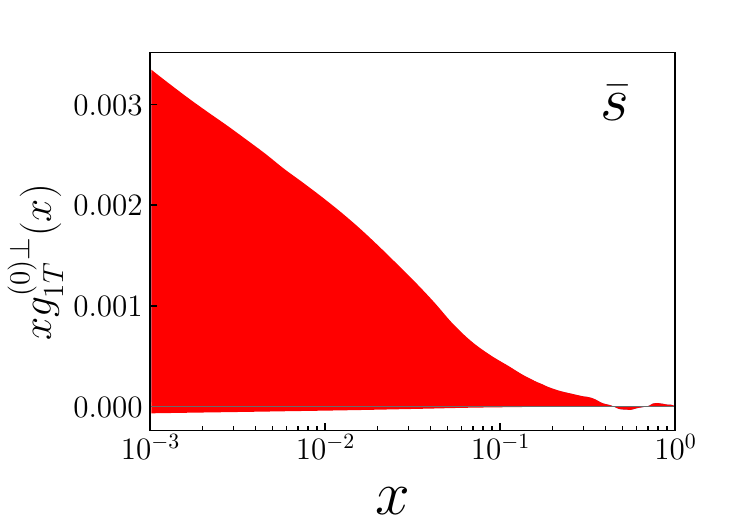}}
  \caption{EicC projections of the zeroth transverse moment of the worm-gear functions, $g_{1T}^{\perp(0)}(x)$ as defined in Eq.~\eqref{equ::g1T0x}, for $\bar u$, $\bar d$, $s$, and $\bar s$ quarks at the scale $Q=2\,\rm GeV$. The uncertainty bands correspond to 68\% CL estimated from the fits to 1000 replicas. The red bands only contain statistical uncertainties, and the blue bands contain both statistical and systematic uncertainties.}
    \label{Fig:xg1Tx_sea_quark}
\end{figure*}

In this study, we achieve total $\chi^2/N = 0.84$ as listed in Table~\ref{tab:fit-A data chi2 result}, together with $\chi^2$ values for each data set. The expectation values and uncertainties of the parameters are summarized in Table~\ref{tab:fit-A parameters}. In Figs.~\ref{Fig:COMPASS hm}--\ref{Fig:JLab}, we show the comparison between the fit results and experimental data, in which the filled points are included in fit while the open points are not. The extracted worm-gear distribution functions $g_{1T}^\perp(x,k_T)$ are shown in Fig.~\ref{Fig:xg1Txkt} at several $x$-slices. As one can observe from the results, the $u$ quark distribution is positive, while the $d$ quark favors a negative distribution though still consistent with zero. This finding qualitatively agrees with the predictions from the quark model~\cite{Pasquini:2008ax,Bacchetta:1999kz,Ji:2002xn,Boffi:2009sh,Bacchetta:2010si}. In addition, we also evaluate the transverse moments of the worm-gear distributions,
\begin{align}
\label{equ::g1T0x}
g_{1 T}^{\perp(0)}(x)&=2\pi \int_0^{k_T^{\rm max}} k_{T} d k_{T}   g_{1 T}^\perp\left(x,  k_{T}\right),\\
\label{equ::g1T1x}
g_{1 T}^{\perp(1)}(x)&=2\pi\int_0^{k_T^{\rm max}} k_{T} d k_{T} \left(\frac{k_{T}^2}{2M^2}\right) g_{1 T}^\perp\left(x,  k_{T}\right),
\end{align}
where the truncation is chosen as $k_{T}^{\rm max} = 1\,\rm GeV$. The results are shown in Figs.~\ref{Fig:xg1Tx}--\ref{Fig:xg1T1x_sea_quark}, which are compatible with Ref.~\cite{Horstmann:2022xkk}.


\begin{table}
\begin{ruledtabular}
\caption{Results of parameters for world data fit. The central values are the average of the results from 1000 replicas, and the uncertainties correspond to $68\%$ CL. The value of $r$ is provided in unit of $\mathrm{GeV}^2$ and the others are dimensionless.}
        \label{tab:fit-A parameters}
\begin{tabular}{ll|ll}
% \hline\hline
Parameter           & Value                 &
Parameter           & Value                 \\\hline
$N_u$               &$0.0206^{+0.0058}_{-0.0050}$      &
$\alpha$            &$16.59^{+65.88}_{-10.11}$      \\
$N_d$               &$-0.0073^{+0.0079}_{-0.0082}$   & 
$\beta$             &$5.57^{+28.65}_{-3.87}$      \\
$10^9r$             &${1.52} ^{+9.25}_{-1.49}$   
\end{tabular}
\end{ruledtabular}
\end{table}

\begin{table}
\caption{The list of results of $\chi^2/N$ of world data fit for each world data set.}
\label{tab:fit-A data chi2 result}
\begin{ruledtabular}
\begin{tabular}{lllllllll}
Data set        & data points   & $\chi^2/N$      \\\hline

HERMES $\pi^{+}$ \cite{HERMES:2020ifk}&          26   &   $1.01 $ \\
HERMES $\pi^{-}$ \cite{HERMES:2020ifk}&          26   &   $0.75 $ \\
HERMES $K^{+}$  \cite{HERMES:2020ifk}&          26   &   $1.09 $ \\
HERMES $K^{-}$  \cite{HERMES:2020ifk}&          26   &   $0.68 $ \\
HERMES $P$      \cite{HERMES:2020ifk}&          30   &   $0.90 $\\ \hline
COMPASS $h^{+}$ \cite{COMPASS:2016led}&          28   &   $0.44 $  \\
COMPASS $h^{-}$ \cite{COMPASS:2016led}&          26   &   $0.65 $  \\ \hline
JLab $\pi^{+}$  \cite{JeffersonLabHallA:2011vwy}&          2    &   $0.61 $\\
JLab $\pi^{-}$  \cite{JeffersonLabHallA:2011vwy}&          2    &   $1.15 $\\ \hline
Total           &          192  &   $0.84$
\end{tabular}
\end{ruledtabular}
\end{table}

\section{EicC projections}
\label{Sec III.2}

The EicC events are generated at the vertex level using the SIDIS Monte Carlo generator, which has been used in previous studies~\cite{Zeng:2022lbo,Zeng:2023nnb}. To select events in the DIS region, we apply the cuts, 
\begin{gather}
    Q^2>1\,\mathrm{GeV}, \quad 0.3<z<0.7,\\
    W>5\,\mathrm{GeV}, \quad W^{\prime}>2\,\mathrm{GeV},
\end{gather}
where $W= \sqrt{(q+P)^2}$ is the invariant mass of the produced, and 
$W^{\prime}=\sqrt{(q+P-P_{h})^2}$ is the missing mass. According to the detection conditions of the designed EicC detector, we further require the scattered electron momentum $P_{e}>0.35\,\mathrm{GeV}$ and the hadron momentum $P_{h}>0.3\,\mathrm{GeV}$.
In the simulation, we take the $3.5\,\mathrm{GeV}$ polarized electron beam with $80\%$ polarization, the $20\,\mathrm{GeV}$ transversely polarized proton beam with $70\%$ polarization, and the $40\,\mathrm{GeV}$ transversely polarized $^{3}\mathrm{He}$ beam with $70\%$ polarization.
Aiming at a complete separation of contributions from all light flavor quarks, we take into account both $\pi^{\pm}$ and $K^{\pm}$ data.

To quantify the impact, we assume $50\,\mathrm{fb}^{-1}$ integrated luminosities of $ep$ and $e^{3}\mathrm{He}$ collisions, which can be achieved with about one-year run according to the proposed instantaneous luminosity. 
For the systematic uncertainties, we assign 2\% relative uncertainty to the polarization of the electron beam, 3\% relative uncertainty to the polarization of the ion beam, and 5\% relative uncertainty to the $^3$He nuclear effect. These are expected the dominant sources of systematic uncertainties based on our current knowledge from existing polarized SIDIS measurements. Because the detailed design of the detectors are still unavailable, we leave more realistic estimation of systematic uncertainties to future studies.

\begin{figure*}[htbp]
  \centering
  {\includegraphics[scale=0.6]{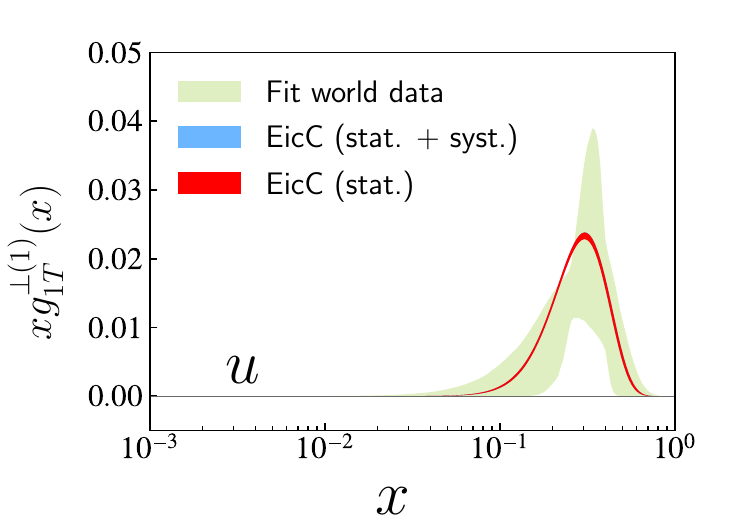}}
  {\includegraphics[scale=0.6]{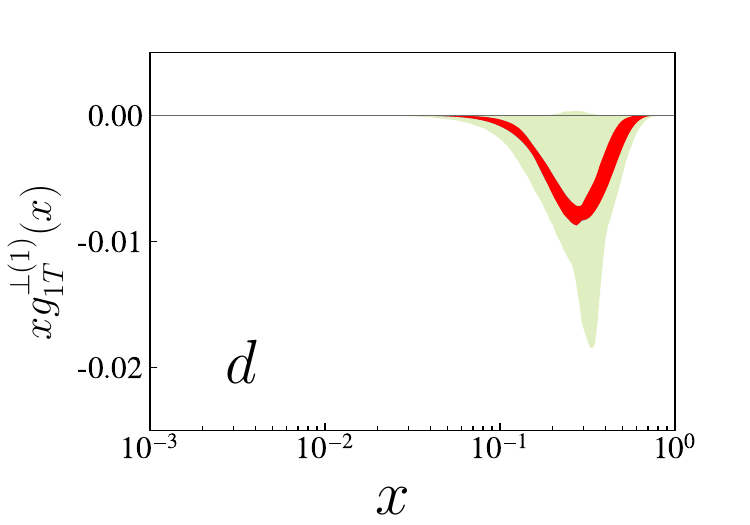}}
  \caption{The first transverse moment of the worm-gear functions, $g_{1T}^{\perp(1)}(x)$ as defined in Eq.~\eqref{equ::g1T1x}, for $u$ and $d$ quarks at the scale $Q=2\,\rm GeV$. The uncertainty bands correspond to 68\% CL estimated from the fits to 1000 replicas. The green bands are extracted distributions by fitting the world SIDIS data, the red bands are EicC projections with only statistical uncertainties, and the blue bands are EicC projections with both statistical and systematic uncertainties.}
    \label{Fig:xg1T1x}
\end{figure*}

\begin{figure*}[htbp]
  \centering
  {\includegraphics[scale=0.55]{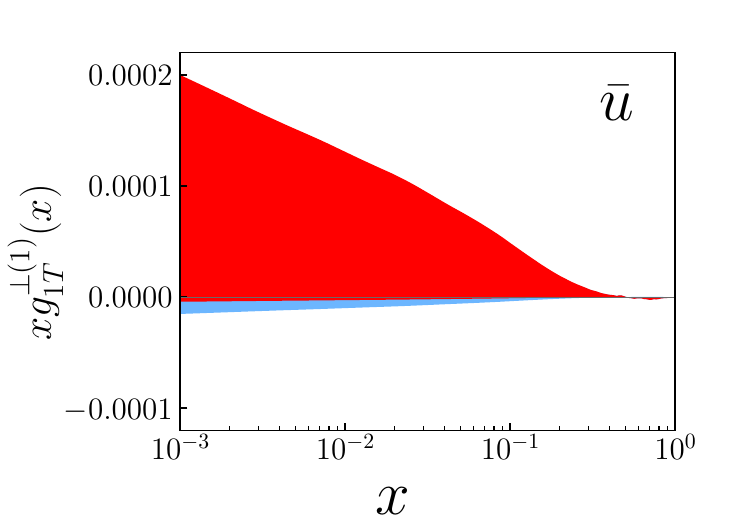}}
  {\includegraphics[scale=0.55]{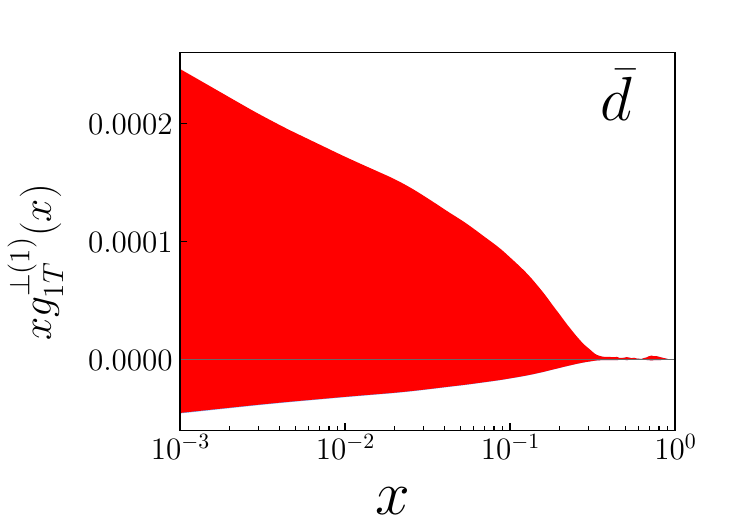}}
  {\includegraphics[scale=0.55]{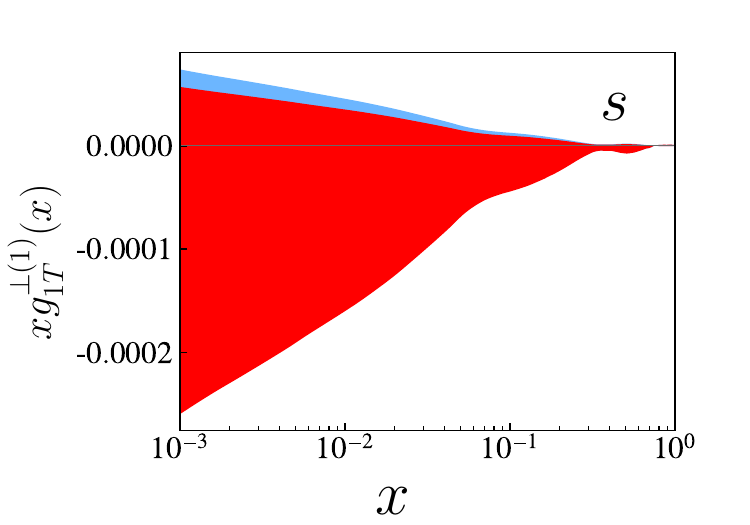}}
  {\includegraphics[scale=0.55]{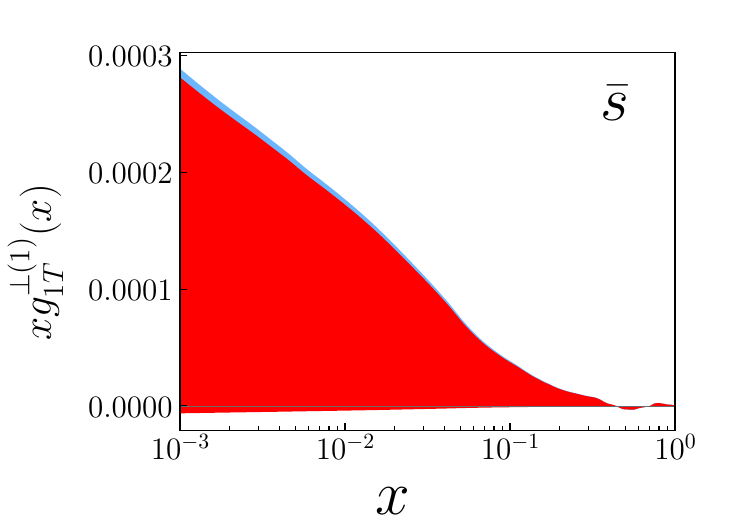}}
  \caption{EicC projections of the first transverse moment of the worm-gear functions, $g_{1T}^{\perp(1)}(x)$ as defined in Eq.~\eqref{equ::g1T1x}, for $\bar u$, $\bar d$, $s$, and $\bar s$ quarks at the scale $Q=2\,\rm GeV$. The uncertainty bands correspond to 68\% CL estimated from the fits to 1000 replicas. The red bands only contain statistical uncertainties, and the blue bands contain both statistical and systematic uncertainties.}
    \label{Fig:xg1T1x_sea_quark}
\end{figure*}

The central values of the worm-gear asymmetry for the EicC pseudodata are evaluated from world data fit, which only include nonvanishing contributions from $u$ and $d$ quarks.
Owing to the considerable amount of the expected EicC data, we can adopt stricter criteria to select data in the TMD region. Hence we set the cut as $\delta < 0.3$, and $5008$ pseudodata points are included. Besides, we also free the parameters $N_q$ for sea quark distributions and choose the same $r_q$ for $\bar u$, $\bar d$, $s$, and $\bar s$ as  
\begin{equation}
    r_s = r_{\bar{u}} = r_{\bar{d}} = r_{\bar{s}} = (r_{u}+r_{d})/2.
\end{equation}
Then there are 12 free parameters in our fit summarized in Table \ref{tab:fit-B parameters}. Following the same procedure, we perform a simultaneous fit to the world data and the EicC pseudodata.
This analysis gives $\chi^2/N = 1.09$ with corresponding values and uncertainties of the parameters listed in Table~\ref{tab:fit-A parameters}, which are evaluated from the fits to 1000 replicas. The EicC projections of the worm-gear distributions $g_{1T}^\perp(x,k_T)$ are shown in Fig.~\ref{Fig:xg1Txkt}. The zeroth transverse momentum moments $g_{1T}^{\perp(0)}(x)$ are shown in Figs.~\ref{Fig:xg1Tx} and~\ref{Fig:xg1Tx_sea_quark}, and the first transverse momentum moments $g_{1T}^{\perp(1)}(x)$ are shown in Figs~\ref{Fig:xg1T1x} and~\ref{Fig:xg1T1x_sea_quark}.

\section{SUMMARY} \label{Sec IV}
In this work, we perform a global fit to the worm-gear asymmetries from SIDIS in a small transverse momentum region, including the TMD evolution effect at the next-to-next-to-leading-logarithmic (NNLL) accuracy.
Due to the fact that the existing experimental uncertainties are too large to determine the worm-gear distributions of sea quarks, only up and down quarks are considered in the global fit.
Then an impact study is performed by including the EicC pseudo data in our global fit. For EicC
pseudo data, the statistical uncertainties and dominant systematic uncertainties are taken into account.
The latter is mainly due to the uncertainties from beam polarimetry and the uncertainties of $^3$He nuclear effects. 

Once the precise data are available from EicC, the precision of the worm-gear distributions for up and down quark will be significantly improved. Meanwhile, it will also provide the opportunity to extract the worm-gear distributions of sea quarks. With much more expected precise data from EicC, one can extract the TMDs utilizing more flexible parametrizations and thus less biased determination of the nucleon spin structures. Owing to the high precision and a wide phase space coverage of EicC pseudo data, a more strict cut of $\delta$, $W$ and $W^{\prime}$ will be feasible. It allows us to have clearer selection of data required by the TMD factorization. On the other hand, the events in the transition region are also valuable to test the matching between TMD and collinear regions.
The combination of polarized $ep$ and $e^{3}\mathrm{He}$ data at similar kinematics are essential for a complete flavor separation.
It is important to remark that the kinematics coverage of EicC will fill the gap between the on-going JLab-$12\,\mathrm{GeV}$ program and the approved Electron-Ion Collider to be built at BNL. With all these facilities, we will be able to have a complete physical picture of nucleon three-dimensional structures, towards a profound understanding of strong interactions.

\begin{table*}
    \begin{ruledtabular}
    \caption{Results of parameters for EicC pseudo data fit. The central values are the average of the results from 1000 replicas, and the uncertainties correspond to $68\%$ CL. The values of $r_{u}$ and $r_{d}$ are provided in unit of $\mathrm{GeV}^2$ and the others are dimensionless.}
            \label{tab:fit-B parameters}
    \begin{tabular}{lll|lll}
    Parameter           & Stat.                             & Stat. + Syst.                         &
    Parameter           & Stat.                             & Stat. + Syst.                         \\\hline
    $N_u$               &$0.0209^{+0.0006}_{-0.0006}$       & $0.0209^{+0.0006}_{-0.0006}$          &
    $\alpha_u$          &$12.46^{+0.72}_{-0.67}$            & $12.47^{+0.67}_{-0.64}$               \\
    $N_d$               &$-0.0077^{+0.0008}_{-0.0009}$      & $-0.0077^{+0.0008}_{-0.0008}$         & 
    $\alpha_d$          &$13.01^{+4.68}_{-2.70}$            & $12.94^{+6.22}_{-2.83}$               \\
    $N_s$               &$-0.00023^{+0.00044}_{-0.00046}$   & $-0.00026^{+0.00042}_{-0.00047}$      &
    $\beta_u$           &$4.46^{+0.26}_{-0.25}$             & $4.46^{+0.23}_{-0.23}$                \\
    $N_{\bar{u}}$       &$0.00019^{+0.00023}_{-0.00022}$    & $0.00020^{+0.00024}_{-0.00021}$       & 
    $\beta_d$\          &$4.31^{+1.47}_{-0.88}$             & $4.29^{+1.80}_{-0.90}$                \\
    $N_{\bar{d}}$       &$0.00021^{+0.00032}_{-0.00031}$    & $0.00022^{+0.00034}_{-0.00032}$       & 
    $r_{u}$             &$0.0067^{+0.0050}_{-0.0048}$       & $0.0067^{+0.0053}_{-0.0050}$          \\
    $N_{\bar{s}}$       &$0.00038^{+0.00045}_{-0.00037}$    & $0.00038^{+0.00045}_{-0.00037}$       & 
    $r_{d}$             &$0.016^{+0.025}_{-0.016}$          & $0.016^{+0.024}_{-0.016}$             \\
    \end{tabular}
    \end{ruledtabular}
    \end{table*}


\acknowledgments{
This work is supported in part by the Strategic Priority Research Program of the Chinese Academy of Sciences under grant number XDB34000000, the Guangdong Major Project of Basic and Applied Basic Research No. 2020B0301030008, the Guangdong Provincial Key Laboratory of Nuclear Science with No. 2019B121203010, the National Natural Science Foundation of China under Grants No.~12175117 and No.~12321005, and Shandong Province Natural Science Foundation Grant No.~ZFJH202303.
P.~Sun is supported by the Natural Science Foundation of China under Grants No.~11975127
and No.~12061131006. 
B.-Q.~Ma is supported by National Natural Science Foundation of China under Grants No.~12075003 and No.~12335006.
The authors also acknowledge the computing resource available at the Southern Nuclear Science Computing Center.
}

\appendix
\section{Fourier transforms for TMDs} \label{App:Fourier}
The Fourier transforms for TMDs are
\begin{equation}
\begin{aligned}
f_1\left(x, k_{T}\right) & = \int \frac{d^2 \mathbf{b}_{T}}{4 \pi^2} e^{i \mathbf{b}_{T} \cdot \mathbf{k}_{T}} f_1(x, b_{T})  \\
& = \int_0^{+\infty} \frac{b_{T} d b_{T}}{2 \pi} J_0\left(b_{T} k_{T}\right) f_1(x, b_{T})  ,
\end{aligned}    
\end{equation}
\begin{equation}
\begin{aligned}
f_1(x, b_{T}) & =\int d^2 \mathbf{k}_{T} e^{-i \mathbf{b}_{T} \cdot \mathbf{k}_{T}} f_1\left(x, k_{T}\right)  \\
& =2 \pi \int_0^{+\infty} k_{T} d k_{T} J_0\left(b_{T} k_{T}\right) f_1\left(x, k_{T}\right)  ,
\end{aligned}    
\end{equation}
\begin{equation}
\begin{aligned}
\frac{k_{T}}{M} g_{1T}^\perp\left(x, k_{T}\right) & = \int \frac{d^2 \mathbf{b}_{T}}{4 \pi^2} e^{i \mathbf{b}_{T} \cdot \mathbf{k}_{T}} (-i {b}_{T} M) g_{1T}^\perp(x, b_{T})  ,\\
g_{1T}^\perp\left(x, k_{T}\right) & = \frac{M^2}{k_{T}} \int_0^{+\infty} \frac{b_{T}^2 d b_{T}}{2 \pi} J_1\left(b_{T} k_{T}\right) g_{1T}^\perp(x, b_{T})  ,
\end{aligned}    
\end{equation}
\begin{equation}
\begin{aligned}
(-i M {b}_{T} ) & g_{1T}^\perp(x, b_{T})  =\int d^2 \mathbf{k}_{T} e^{-i \mathbf{b}_{T} \cdot \mathbf{k}_{T}} \frac{k_T}{M} g_{1T}^\perp\left(x, k_{T}\right)  ,\\
g_{1T}^\perp(x, b_{T}) & =\frac{2 \pi }{M^2 b_{T}}\int_0^{+\infty} k_{T}^2 d k_{T} J_1\left(b_{T} k_{T}\right) g_{1T}^\perp\left(x, k_{T}\right)  ,
\end{aligned}    
\end{equation}
\begin{equation}
\begin{aligned}
D_1\left(z, k_{T}\right) & = \int \frac{d^2 \mathbf{b}_{T}}{4 \pi^2} e^{- i \mathbf{b}_{T} \cdot \mathbf{k}_{T}} D_1(z, b_{T})  \\
& = \int_0^{+\infty} \frac{b_{T} d b_{T}}{2 \pi} J_0\left(b_{T} k_{T}\right) D_1(z, b_{T})  ,
\end{aligned}    
\end{equation}
\begin{equation}
\begin{aligned}
D_1(z, b_{T}) & =\int d^2 \mathbf{k}_{T} e^{i \mathbf{b}_{T} \cdot \mathbf{k}_{T}} D_1\left(z, k_{T}\right)  \\
& =2 \pi \int_0^{+\infty} k_{T} d k_{T} J_0\left(b_{T} k_{T}\right) D_1\left(z, k_{T}\right)  .
\end{aligned}    
\end{equation}

\section{Expressions for energy evolution factor} \label{App:Evolution}
The $\mathcal{D}(Q,b_{T})$ is the rapidity anomalous dimension (RAD). At large values of $b_{T}$, the $\mathcal{D}(Q,b_{T})$ behaves like a linear function of $b_{T}$, which is suggested by some models such as \cite{Tafat:2001in,Vladimirov:2020umg}. Therefore, we parameterize the RAD as 
\begin{equation}
\mathcal{D}(\mu, b_{T})=\mathcal{D}_{\text {resum}}\left(\mu, b_{T}^*(b_{T})\right)+c_0 b_{T} b_{T}^*(b_{T}),
\end{equation}
where the $\mathcal{D}_{\text {resum}}\left(\mu, b_{T}^*(b_{T})\right)$ is the resummed perturbative expansion of RAD, and $b_{T}^*(b_{T})$ take the form
\begin{equation}
b_{T}^*(b_{T})=\frac{b_{T}}{\sqrt{1+b_{T}^2 / B_{\mathrm{NP}}^2}}.
\end{equation}
At small values of $b_{T}$, the term $c_0 b_{T} b_{T}^*(b_{T})$ can be ignored and the term $\mathcal{D}_{\text {resum}}$ is dominant, while at large values of $b_{T}$, the $\mathcal{D}$ behave like $c_0 B_{\text{NP}} b_{T}$. 
We take $B_{\text{NP}} = 1.93\,\mathrm{GeV}^{-1}$ and $c_0 = 0.0391\,\mathrm{GeV}^{2}$ as determined in SV19 model \cite{Scimemi:2019cmh}.
The $\mathcal{D}_{\text {resum}}$ can be represented as: 
\begin{equation}
\begin{aligned}
& \mathcal{D}_{\text {resum}}(\mu, b_{T})=-\frac{\Gamma_0}{2 \beta_0} \ln (1-X) \\
& +\frac{a_s}{2 \beta_0(1-X)}\left[-\frac{\beta_1 \Gamma_0}{\beta_0}(\ln (1-X)+X)+\Gamma_1 X\right] \\
& +\frac{a_s^2}{(1-X)^2}\left[\frac{\Gamma_0 \beta_1^2}{4 \beta_0^3}\left(\ln ^2(1-X)-X^2\right)\right. \\
& +\frac{\beta_1 \Gamma_1}{4 \beta_0^2}\left(X^2-2 X-2 \ln (1-X)\right) \\
& +\frac{\Gamma_0 \beta_2}{4 \beta_0^2} X^2-\frac{\Gamma_2}{4 \beta_0} X(X-2) \\
& \left.+C_F C_A\left(\frac{404}{27}-14 \zeta_3\right)-\frac{112}{27} T_R N_f C_F\right],
\end{aligned}
\end{equation}
where $X=\beta_0 a_s \mathbf{L}_\mu$, $\beta_{i}$ are coefficients of anomaly dimension of strong coupling constant, which satisfies
\begin{equation}
\mu^2 \frac{d a_s(\mu)}{d \mu^2}=-\beta\left(a_s\right)=-\sum_{i=0}^{\infty} a_s^{i+2}(\mu) \beta_i.
\end{equation}
$C_A = 3$ and $T_{R} = 1/2$ are color factors of the $SU(3)$.
The $\Gamma_i$ are coefficients of expansion of CUSP anomaly dimension $\Gamma_{\text{cusp}}(\mu)$, which is related with the integrability condition (\ref{Equ:integrability condition}) of the evolution equation.
The $\Gamma_{i}$ are defined by
\begin{equation}
\Gamma_{\text {cusp }}(\mu)=\sum_{i=0}^{\infty} a_s^{i+1} \Gamma_i.
\end{equation}
With the CUSP anomaly dimension, the anomaly dimension $\gamma_{V}$ can be written as
\begin{equation}
\gamma_F(\mu, \zeta)=\Gamma_{\text {cusp }}(\mu) \ln \left(\frac{\mu^2}{\zeta}\right)-\gamma_V(\mu),
\end{equation}
and the $\gamma_V(\mu)$ can be expanded as 
\begin{equation}
\gamma_V(\mu)=\sum_{i=1}^{\infty} a_s^n \gamma_i.
\end{equation}
The $\Gamma_i$ and $\gamma_i$ can be determined perturbatively, and up to two-loop level
\begin{equation}
\begin{aligned}
\Gamma_0 & =4 C_F, \\
\Gamma_1 & =4 C_F\left[\left(\frac{67}{9}-\frac{\pi^2}{3}\right) C_A-\frac{20}{9} T_R N_f\right], \\
\gamma_1 & =-6 C_F, \\
\gamma_2 & =C_F^2\left(-3+4 \pi^2-48 \zeta_3\right) \\
& +C_F C_A\left(-\frac{961}{27}-\frac{11 \pi^2}{3}+52 \zeta_3\right) \\
& +C_F T_R N_f\left(\frac{260}{27}+\frac{4 \pi^2}{3}\right),
\end{aligned}
\end{equation}
where $N_f$ is the number of active quark flavors and have different values at different energy scales (see Table \ref{tab:values of N_f}), $\zeta_3 \approx 1.202$ is the Apéry's constant.

\begin{table}
	\caption{Values of $N_f$ at different values of energy scale.}
 \label{tab:values of N_f}
\begin{tabular}{cc}
\hline\hline
  $\mu \leq 1.27\, \mathrm{GeV}$      & $N_f = 3$ \\
$1.27 < \mu \leq 4.18 \, \mathrm{GeV}$     &   $N_f = 4$  \\
   $\mu>4.18\, \mathrm{GeV}$      & $N_f = 5$ \\
\hline\hline
\end{tabular}
\end{table}

Due to that the non-perturbative corrections to the RAD can not be ignored at large-$b_{T}$, we need to use the exact solution of $\zeta_{\mu}$ at large-$b_{T}$; while at very small-$b_{T}$, we use the perturbative solution. In order to connect these two region, we introduce a $e^{-b_{T}^2/B_{\mathrm{NP}}^2}$ factor, and the $\zeta_{\mu}$ is expressed as \cite{Scimemi:2019cmh}
\begin{equation}
\zeta_\mu(b_{T})=\zeta_\mu^{\text {pert }}(b_{T}) e^{-\frac{b_{T}^2}{B_{\mathrm{NP}}^2}}+\zeta_\mu^{\operatorname{exact}}(b_{T})\left(1-e^{-\frac{b_{T}^2}{B_{\mathrm{NP}}^2}}\right).
\end{equation}
Therefore, at $b_{T}^2 \ll B^2_{\mathrm{NP}}$, $\zeta_{\mu}$ is dominantly given by perturbative solution, and at other regions, it will turn to exact solution.
We express the $\zeta_\mu^{\text {pert }}$ and $\zeta_\mu^{\operatorname{exact}}$ here,
\begin{equation}
\zeta_\mu^{\text {pert }}(\mu, b_{T})=\frac{2 \mu e^{-\gamma_E}}{b_{T}} e^{-v(\mu, b_{T})},
\end{equation}
\begin{equation}
\zeta_\mu^{\operatorname{exact}}(\mu, b_{T})=\mu^2 e^{-g(\mu, b_{T}) / \mathcal{D}(\mu, b_{T})},
\end{equation}
where
\begin{equation}
v(\mu, b_{T})=\frac{\gamma_1}{\Gamma_0}+a_s\left[\frac{\beta_0}{12} \mathbf{L}_\mu^2+\frac{\gamma_2+d_2(0)}{\Gamma_0}-\frac{\gamma_1 \Gamma_1}{\Gamma_0^2}\right],
\end{equation}
and 
\begin{widetext}
\begin{equation}
\begin{aligned}
& g(\mu, b_{T})=\frac{1}{a_s} \frac{\Gamma_0}{2 \beta_0^2}\left\{e^{-p}-1+p+ 
 a_s\left[\frac{\beta_1}{\beta_0}\left(e^{-p}-1+p-\frac{p^2}{2}\right)\right. \left.-\frac{\Gamma_1}{\Gamma_0}\left(e^{-p}-1+p\right)+\frac{\beta_0 \gamma_1}{\Gamma_0} p\right]\right.+\\
& a_s^2\left[\left(\frac{\Gamma_1^2}{\Gamma_0^2}-\frac{\Gamma_2}{\Gamma_0}\right)(\cosh p-1) \left.+\left(\frac{\beta_1 \Gamma_1}{\beta_0 \Gamma_0}-\frac{\beta_2}{\beta_0}\right)(\sinh p-p)+\left(\frac{\beta_0 \gamma_2}{\Gamma_0}-\frac{\beta_0 \gamma_1 \Gamma_1}{\Gamma_0^2}\right)\left(e^p-1\right)\right]\right\}.
\end{aligned}
\end{equation}
\end{widetext}
In the $g(\mu,b_{T})$, the $p$ is 
\begin{equation}
p=\frac{2 \beta_0 \mathcal{D}(\mu, b_{T})}{\Gamma_0},
\end{equation}
and in the $v(\mu, b_{T})$, the $d_{2}(0)$ is 
\begin{equation}
d_2(0)=C_F C_A\left(\frac{404}{27}-14 \zeta_3\right)-\frac{112}{27} T_R N_f C_F.
\end{equation}

\end{document}